\documentclass[twocolumn]{aastex701}

\usepackage{makecell}
\usepackage{longtable}
\usepackage{hyperref}
\usepackage{amsmath}
\usepackage{dsfont}

\newcommand{\customref}[2]{\hyperref[#2]{#1}}
\defcitealias{gellerWIYNOpenCluster2008}{G08}
\defcitealias{geller_wiyn_2009}{G09}
\defcitealias{geller_wiyn_2012}{G12}

\begin{document}

\title{WIYN Open Cluster Study. XCVII. An Extended Radial-Velocity Survey and Spectroscopic Binary Orbits in the Open Cluster NGC 188}
\shorttitle{WOCS XCVII: A Renewed Radial Velocity Study in NGC 188}
\shortauthors{Narayan et al.}

\author[0009-0007-0488-5685]{Ritvik Sai Narayan}
\affiliation{Department of Astronomy, University of Wisconsin-Madison, 475 N. Charter St., Madison, WI 53706, USA}
\email[show]{rnarayan4@wisc.edu}

\author[0009-0006-7474-7889]{Evan Linck}
\affiliation{Department of Astronomy, University of Wisconsin-Madison, 475 N. Charter St., Madison, WI 53706, USA}
\email{elinck@wisc.edu}

\author[0000-0002-7130-2757]{Robert D. Mathieu}
\affiliation{Department of Astronomy, University of Wisconsin-Madison, 475 N. Charter St., Madison, WI 53706, USA}
\email[show]{mathieu@astro.wisc.edu}

\author[0000-0002-3881-9332]{Aaron M. Geller}
\affiliation{Center for Interdisciplinary Exploration and Research in Astrophysics (CIERA) and Department of Physics and Astronomy, Northwestern University, 1800 Sherman
Ave, Evanston, IL 60201, USA}
\email{a-geller@northwestern.edu}

\begin{abstract}
We present 35 new spectroscopic-binary orbits from our extended radial-velocity (RV) survey of the old ($6.4 \pm 0.2$ Gyr) open cluster NGC 188. Using data from the WIYN Open Cluster Study (WOCS) and APOGEE-2, this work nearly doubles the temporal baseline of the previous RV study of NGC 188. We obtain orbital solutions within a stellar sample that spans a magnitude range of {$10.8 \leq \mathrm{G} \leq 16.5 \; (0.9\mbox{-}1.2 \; {M_\odot})$}. With revised membership determinations using Gaia DR3 proper-motions and parallaxes, we reassess the cluster binary frequency and period-eccentricity distribution. The incompleteness-corrected binary frequency is $33.1 \% \pm 3.8\%$ for periods less than $10^4$ days, and the tidal-circularization period is $14.4^{+0.14}_{-0.11}$ days. We find evidence that giants are deficient in short-period orbits, and suggest that the missing giants may have undergone mass transfer and in part formed the population of blue straggler stars (BSSs) and blue lurkers. Among the binaries of note, we highlight WOCS 3953 as a blue lurker candidate, WOCS 5020 and WOCS 4945 as very long-period eccentric BSSs, and WOCS 4230, a BSS with a very close WD companion. 
\end{abstract}

\keywords{\uat{Binary stars}{154} --- \uat{Spectroscopic binary stars}{1557} --- \uat{Open star clusters}{1160}}

\section{Introduction} \label{sec:intro}
Open star clusters provide crucial insights into the evolution of stellar populations over time, forming natural laboratories given their coeval formation. Interactions between stars and their companions in these clusters can significantly impact their evolution, leading to the formation of stars that are not explained by single-star stellar evolution models, such as blue straggler stars (BSSs) and blue lurkers (BLs). 

NGC 188 is a well-studied, rich, and old open cluster that has been a focus of extensive research since the 1960s. Located at $(l,b)=(122.8,+22.4^\circ)$, this cluster lies above the Galactic disk at a distance of 1852 $\pm$ 76 pc \citep{deniz_sed_2024} with a metallicity of [Fe/H] = $-$0.03 $\pm$ 0.015 dex \citep{casamiquela_abundanceage_2021} and an estimated age ranging between 5.78 Gyr and 7.65 Gyr \citep{hills_bayesian_2015, childs_goodbye_2024,meibom_age_2009,yakut_novel_2025, friel_abundances_2010,deniz_sed_2024}. Furthermore, the well-defined morphological structure of the NGC 188 main-sequence population allows for the relatively clear identification of BSSs, enabling detailed spectroscopic binary studies that reveal a high BSS binary fraction \citep{mathieu_binary_2009} and strong evidence for mass transfer origins \citep{gosnell_implications_2015,gosnell_constraining_2019,subramaniam_hot_2016,rain_binary_2024}. Given our long temporal coverage of the cluster, its extensively characterized binary population, and its old age, NGC 188 stands out as an attractive open cluster for studying binary evolution and assessing alignments between theory and observation. 

Over the decades, numerous observational campaigns have been conducted to determine the cluster membership and properties. One of the recent comprehensive studies has been that of the WIYN Open Cluster Survey \citep[WOCS;][]{mathieu_wiyn_2000}, which has provided an extensive dataset of ground-based proper motions \citep{platais_wocs_2003}, abundances \citep{milliman_phd_2016, sun_wiyn_2022}, optical and time-series photometric data \citep{frinchaboy_wiyn_2008}, and RVs (\citealt{gellerWIYNOpenCluster2008, geller_wiyn_2009,geller_wiyn_2012}, hereafter \citetalias{gellerWIYNOpenCluster2008}, \citetalias{geller_wiyn_2009}, and \citetalias{geller_wiyn_2012}) for NGC 188. This work is the fourth paper in the series of WOCS RV studies of the NGC 188 field. 

\setcounter{footnote}{0}
These studies have utilized the WIYN\footnote{The WIYN 3.5m Observatory is a joint facility of the University of Wisconsin–Madison, Indiana University, NSF’s NOIRLab, the Pennsylvania State University, and Princeton University.} Hydra Multi-Object Spectrograph (MOS) to obtain high-precision RV measurements and thereby improve cluster membership determinations, identify numerous binary systems, provide orbital solutions, measure the tidal-circularization period \citep{mathieu_wiyn_2004}, yield determinations of the cluster binary fraction, and apply $N$-body modeling to understand the formation and evolution history of the observed binary and BSS populations \citep{geller_direct_2013}. 

With 16 additional years of WOCS data (2008-2024) and combination with RV data from APOGEE-2 \citep[2011-2020]{abdurrouf_seventeenth_2022}, this work nearly doubles the time baseline of the previous WIYN observations (1996-2008), allowing for a comprehensive reassessment of the binary population. The extended temporal coverage enhances our ability to detect long-period binary systems and refine the orbital parameters of previously identified binaries. Furthermore, the advent of precise astrometry from the \textit{Gaia Space Observatory} allows more accurate proper-motion membership determination, significantly improving the identification of cluster members.

This paper is organized as follows: Section~\ref{sec:photometry} details our stellar sample and proper-motion membership determinations. In Section~\ref{sec:data}, we present our complete RV database. Section~\ref{sec:membership} discusses single- and binary-star membership probabilities and presents our isochrone fit to the cluster members. Section~\ref{sec:orbits} introduces our methodology for deriving binary orbits and discusses the solutions for 35 single-lined (SB1) binary cluster members obtained from this survey. Section~\ref{sec:discussion} revisits tidal circularization in this cluster with our larger binary sample, discusses our findings on mass transfer in NGC 188, and notable binaries, including a potential BL candidate. Finally, Section~\ref{sec:summary} gives a brief summary. An \customref{Appendix}{sec:appendix} provides the orbital solutions for 44 SB1 field binaries discovered serendipitously in the NGC 188 field during this survey.

\section{Stellar Sample} \label{sec:photometry}

The WOCS RV study of stars in the NGC 188 field has gone through three stages of its target sample. The initial sample in 1996 comprised 1127 stars with $\mathrm{B} < 16.2$ from the proper-motion study of \citet{dinescu_proper-motion_1996}. Prior to the publication of \citetalias{gellerWIYNOpenCluster2008}, the sample was rebuilt to include stars in the proper-motion study of \citet{platais_wocs_2003}, which contained proper-motion member stars with $\mathrm{V} < 21$ within $0.5^{\circ}$ of the cluster center. In the first WOCS papers, the sample included 1498 stars in the NGC 188 field with $\mathrm{V} < 16.5$ and within $0.5^{\circ}$. Proper-motion membership probability was used to set the priority of RV observations, but the entire sample included both proper-motion members and non-members.

\begin{figure}[h!]
    \centering
    \includegraphics[width=\linewidth]{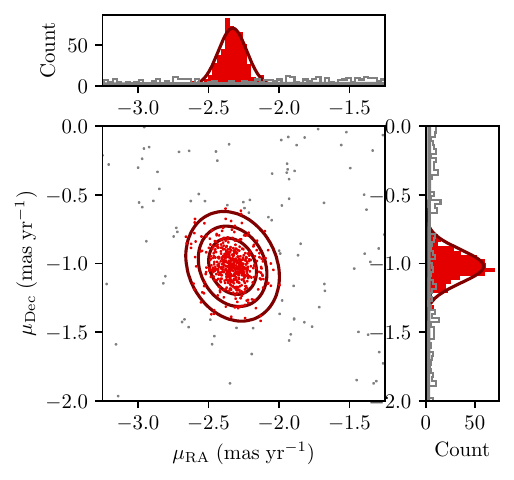}
    \caption{The two-dimensional proper motions of stars with magnitudes  G $\leq$ 16.5  in the field of NGC 188 are displayed. Proper-motion members of NGC 188 are shown in red, while field stars are depicted in gray. Contours in the main panel represent the  $1\sigma$,  $2\sigma$, and  $3\sigma$ levels of the Gaussian distribution for the cluster, along with the  $0.5\sigma$  level for the field’s Gaussian distribution. The top and side panels present the one-dimensional projections of the stellar and Gaussian distributions for the cluster and the field, the latter plotted in red and gray, respectively. }
    \label{fig:pmm_analysis}
\end{figure}

This study constructs a third stellar sample using proper-motion and parallax information from Gaia DR3 \citep{vallenari_gaia_2023}. Given the 1$^\circ$ field of view of the WIYN Hydra MOS and the integration time providing the numerous RV observations for binary orbit solutions, we limited our search for potential cluster members to stars within $0.5^{\circ}$ of the cluster center and $\mathrm{G} \leq 16.5$ mag. Following the procedure of \cite{linck_wiyn_2024}, we performed a Gaussian mixture model (GMM) analysis of the proper-motion distributions of the cluster and the field (see Figure~\ref{fig:pmm_analysis}). For this third target sample, we required stars to be proper-motion members, i.e., have a proper-motion membership probability greater than 50\% ({551} stars total). 

Table~\ref{table:pm_gmm_table} presents the parameters of our best-fit Gaussian distributions for cluster and field proper motions. Finally, we required the error on parallax for each star to overlap with the parallax distribution of proper-motion members at the 3$\sigma$ level, which removed 5 stars, leading to a final sample of {546} stars. 

Table~\ref{table:star_summary} presents summarized information for each of the 546 stars in this study, as well as for 707 stars for which we have previously made RV observations but are not Gaia DR3 proper-motion members, listing a total of 1253 stars. The first row includes the WOCS ID (identification system comes from \citealt{platais_wocs_2003} and is referred to as PKM previously in the series), the Gaia DR3 source ID, right ascension ($\alpha$), declination ($\delta$), Gaia G magnitude, Gaia color ($\mathrm{G_{BP} - G_\mathrm{RP}}$), number of observations from each of WOCS and APOGEE ($\mathrm{N_W \; \&\;N_A}$), mean RV ($\mathrm{\overline{RV}}$), standard-deviation of RV ($e$), internal error ($i$), median $v \sin i$ measurement, $e/i$, proper-motion membership probability from the GMM analysis ($\mathrm{P_\mu}$), RV membership probability ($\mathrm{P_{RV}}$), classification (see Section~\ref{sec:membership}), a Boolean variable that indicates whether the star is within the cluster's parallax cutoff ($\mathds{1} (\pi_c)$), and comments. This table includes proper-motion members for which we do not have RVs for completeness of the sample. These stars are classified as U and do not have anything listed for values derived from RV measurements (e.g., $\mathrm{\overline{RV}}$,  $v \sin i $, $\mathrm{P_{RV}}$).

\begin{deluxetable}{lcr}
\tablehead{\colhead{Parameter} &\colhead{Cluster}
&\colhead{Field}}
\tabletypesize{\footnotesize}
\tablewidth{1.0\columnwidth}
\tablecaption{Cluster and Field Gaussian Distributions of Proper Motion\label{table:pm_gmm_table}}
\startdata
     $\mathrm{Ampl._{RA}\; (count)}$ & $71.22 \pm 1.00 $ & $3.19 \pm 0.01 $ \\
     $\mathrm{\overline{\mu}_{RA}\;(mas\;yr^{-1})}$ & $-2.328 \pm  0.001$ & $-0.926\pm 0.003 $ \\
     $\mathrm{Ampl._{Dec}\; (count)}$ & $59.07 \pm 0.86 $ & $3.94 \pm 0.01 $ \\
     $\mathrm{\overline{\mu}_{Dec}\;(mas\;yr^{-1})}$ & $-1.021 \pm  0.002$ & $0.717 \pm 0.004 $ \\
     $\mathrm{\sigma_{RA} \;(mas\;yr^{-1})}$ & $0.0103 \pm 0.0004 $ & $ 15.994 \pm 0.029 $ \\
     $\mathrm{\sigma_{Dec} \;(mas\;yr^{-1})}$ & $0.0149 \pm 0.0005 $ & $ 10.245\pm  0.019 $ \\
\enddata
\end{deluxetable}

\begin{deluxetable*}{l c c c c c c c c c c c c c c c c p{2cm}}
\tabletypesize{\scriptsize}
\tablewidth{0pt}
\tablecaption{Star Summary Table\label{memberships}}
\tablehead{\colhead{ID} & \colhead{Gaia DR3} &\colhead{$\mathrm{\alpha}$} & \colhead{$\mathrm{\delta}$} & \colhead{$\mathrm{G}$} & \colhead{$\mathrm{G_{RP} - G_{BP}}$} & \colhead{$\mathrm{N_{W}}$} & \colhead{$\mathrm{N_{A}}$} & \colhead{$\mathrm{\overline{RV}}$} & \colhead{$e$} &\colhead{$i$} & \colhead{$v\sin i$}& \colhead{$e/i$} &  \colhead{$\mathrm{P_{\mu}}$}&\colhead{$\mathrm{P_{RV}}$} &  \colhead{$\mathrm{Class}$}& \colhead{${\mathds{1}(\pi_{c})}$} & \colhead{Comments}
} \label{table:star_summary}
\startdata
1007 & 573980906906588928 & 10.725079 & 85.509554 & 16.17 & 1.26 & 2 & 0 & -58.39 & 0.41 & 0.4 & $<10$ & 1.02 & 0.0 & 0.0 & UNM & False &  \\
1013 & 573980498886619648 & 10.507378 & 85.493812 & 15.58 & 0.85 & 10 & 0 & -56.92 & 0.51 & 0.4 & $<10$ & 1.27 & 0.0 & 0.0 & SNM & True & \\
1025 & 573968679136630528 & 11.04614 & 85.463065 & 15.09 & 0.85 & 9 & 0 & -42.26 & 0.24 & 0.4 & $<10$ & 0.61 & 1.0 & 0.97 & SM & True & \\
3037 & 573573885744009984 & 7.961516 & 85.13324 & 10.06 & 2.18 & 0 & 0 & -42.16 & 0.14 & - & - & - & 0.97 & 0.97 & SM & True & The reported RV mean and std are from Gaia DR3.\\
4230 & 573944283722429952 & 10.849436 & 85.342409 & 14.99 & 0.76 & 31 & 0 & -38.55 & 34.22 & 1.65 & 106 & 20.74 & 1.0 & 0.0 & BU & True & \\
6407 & 573782724232070912 & 15.227895 & 85.403621 & 14.39 & 0.84 & 25 & 0 & -48.08 & 16.55 & 1.3 & 79 & 12.5 & 0.0 & 0.0 & BNM & True & Orbit solution in Geller+2009. \\
8406 & 573932670130981632 & 11.43724 & 85.071728 & 14.53 & 1.2 & 32 & 36 & -42.2 & 1.09 & 0.4 & $<10$ & 2.74 & 0.99 & 0.97 & BM & True & \\
8894 & 573929749553233664 & 12.873472 & 84.997225 & 15.14 & 0.92 & 25 & 0 & -42.11 & 30.66 & 0.4 & $<10$ & 76.64 & 1.0 & 0.97 & BM & True & \\
\enddata
\tablecomments{This table is available in machine-readable format. $\mathrm{\alpha\;and\;\delta}$ are at epoch J2015.5. G and $\mathrm{G_{RP} - G_{BP}}$ represent Gaia G magnitudes and color. $\mathrm{N_W \; and \; N_A}$ are number of observations from WOCS and APOGEE respectively. $\mathrm{\overline{RV}},\;e,\;i,\;\mathrm{and}\;v\sin i$ are in $\mathrm{km\;s^{-1}}$. $\mathrm{P_{\mu}}$ is the probability of proper-motion membership from the GMM analysis, while $\mathrm{P_{RV}}$ is the probability of RV membership. Class represents the type of membership assigned and $\mathds{1}(\pi_c)$ is an indicator variable for whether the star is within the cluster's parallax or not. Lastly, we also show comments.}
\end{deluxetable*}

\section{Radial-Velocity Data} \label{sec:data}

\subsection{Overview} \label{subsec:RVoverview}

Our extended RV database adds two new datasets, one from WIYN Hydra and one from the APOGEE-2 DR17 mission of the Sloan Digital Sky Survey (SDSS) \citep{abdurrouf_seventeenth_2022}. The latter data release focused on surveying giants in the NGC 188 color-magnitude space, allowing us to determine more orbital solutions for the giant population. 

10,567 RV measurements were made at the WIYN 3.5m telescope between 1995 October 30 and 2023 March 17. Our previous RV data release \citepalias{geller_wiyn_2009} contained 8751 measurements through 2008 September 15. This paper adds an additional 1816 RV measurements taken for long-period binary monitoring over roughly three epochs: 2008--2009 (824 observations), 2012--2017 (399 observations), and 2020--2023 (593 observations).  

Observations of 157 stars in the NGC 188 field from the APOGEE-2 survey were taken between 11 September 2011 and 6 November 2020. Table~\ref{table:rvdata} also includes these RV measurements, corrected with the zero-point offset we find with respect to WOCS data (Section~\ref{subsec:zeropoint}). These measurements do not provide a CCF peak height.

We next provide brief summaries of WIYN and APOGEE instrumentation, data reduction processes, and measurement precision in this section for convenience. For more details, refer to \citetalias{gellerWIYNOpenCluster2008} and \cite{majewski_apache_2017}, respectively. 

Our database also contains early-epoch RV data from the Dominion Astrophysical Observatory (DAO) and the Palomar Observatory. These RVs were measured at the DAO 1.2m telescope (1980-1996; 499 measurements) and the Palomar 5m telescopes (1973-1980; 77 measurements). In total, these measurements extended the \citetalias{gellerWIYNOpenCluster2008} stellar sample's bright limit to $V = 10.8$ \textemdash bringing in evolved subgiants, giants, and BSSs \textemdash and for select targets extended the time baselines to 35 years in 2008. 

In Table~\ref{table:rvdata}, we present the RV data for 1204 stars in Table~\ref{table:star_summary} for which we made RV measurements. We show the first entry here and provide the entire table electronically. For individual RV measurements, we provide the reduced Heliocentric Julian Date ($\mathrm{HJD} - 2,400,000$), the measured RV, the cross-correlation function height (with a maximum of 1), the residuals ($O-C$) and phases for those stars with orbital solutions, and the observatory at which the observations were taken, using ``W'' for WIYN, ``A'' for APOGEE,``D'' for DAO, and ``P'' for Palomar. For previously determined double-lined binaries (SB2s) with orbital solutions, we provide RVs and cross-correlation peak heights (where available) for both stars and their respective residuals.

We note that Table~\ref{table:rvdata} includes field stars for which we previously made RV measurements, but that we do not consider proper-motion members and are not part of the current analyses. 

\begin{deluxetable*}{cccccccccccc} \label{table:rvdata}
\tablewidth{0pt}
\tabletypesize{\footnotesize}
\tablecaption{Radial Velocity Measurements}

\tablehead{\colhead{ID} & \colhead{HJD - 2,400,000} & \colhead{Telescope} & \colhead{$\mathrm{RV_1}$} & \colhead{Correlation} & \colhead{$\mathrm{O-C_1}$} & \colhead{$\mathrm{RV_2}$} & \colhead{Correlation} & \colhead{$\mathrm{O-C_2}$} & \colhead{phase} \\ 
\colhead{} & \colhead{(days)} & \colhead{} & \colhead{$\mathrm{(km\,s^{-1})}$} & \colhead{$\mathrm{Height_1}$} & \colhead{$\mathrm{(km\,s^{-1})}$} & \colhead{$\mathrm{(km\,s^{-1})}$} & \colhead{$\mathrm{Height_2}$} & \colhead{$\mathrm{(km\,s^{-1})}$} & \colhead{} }

\startdata
6125 & 2453390.018 & W & -44.00 & 0.74 & -0.2 & - & - & - & 0.91 \\
6125 & 2453722.762 & W & -43.70 & 0.72 & -0.4 & - & - & - & 0.10 \\
6125 & 2454279.695 & W & -41.70 & 0.41 & 0.3 & - & - & - & 0.13 \\
6125 & 2454751.796 & W & -36.87 & 0.49 & -0.3 & - & - & - & 0.32 \\
6125 & 2454904.937 & W & -36.32 & 0.47 & -0.4 & - & - & - & 0.63 \\
6125 & 2459210.698 & W & -35.74 & 0.44 & 0.0 & - & - & - & 0.61 \\
6125 & 2459271.918 & W & -44.61 & 0.91 & 0.1 & - & - & - & 0.94 \\
6125 & 2459275.907 & W & -45.71 & 0.55 & 0.2 & - & - & - & 0.02 \\
6125 & 2459630.859 & W & -36.30 & 0.90 & 0.4 & - & - & - & 0.69 \\
6125 & 2459890.769 & W & -36.54 & 0.62 & 0.4 & - & - & - & 0.30 \\
\enddata
\tablecomments{This table is available in machine-readable format. A portion is shown here for guidance regarding its form and content.}
\end{deluxetable*}

\subsection{WIYN Data Acquisition} \label{subsec:wiyn_data}

We used the Hydra MOS on the WIYN 3.5m telescope at Kitt Peak, Arizona to collect spectra in the NGC 188 field. The telescope has a $1^\circ$-diameter field of view, with 0.8' median seeing. In one ``pointing" the MOS can obtain simultaneous spectra of $\sim 80$ stars over the field, with each fiber having a 0.31'' aperture. The spectra are centered on $5130 \; \mathring{\text{A}}$, extending $250 \; \mathring{\text{A}}$ in either direction, allowing us to capture a rich set of absorption lines including the Mg B triplet. Using an echelle grating at 11th order, we achieve a dispersion of $0.13 \; \mathring{\text{A}}$ per pixel and a resolution of $\approx$ 20,000. Typical integration times are 3600 s and 7200 s.

\subsection{WIYN Data Reduction} \label{subsec:wiyn_reduction}
During each pointing, we assign $\sim 70$ fibers to stars, while the remaining are assigned to sky positions for background sky subtraction. From these 70 fibers, our data reduction routine is designed to extract science spectra from CCD images and derive their heliocentric RVs. The process begins with bias subtraction, where a cubic-spline function is fit to the readout bias level and subtracted from each column of the science image. Spectra extraction involves tracing apertures using dome flats, fitting a Legendre polynomial to each spectrum's position, and correcting for fiber-to-fiber throughput variations using flat-fielding techniques. Wavelength calibration is achieved with ThAr spectra taken before and after each science integration, fitting cubic-spline functions to emission-line wavelengths and pixel centroids. The L.A. Cosmic routine \citep{dokkum_cosmicray_2001} was added in 2014 to reject cosmic rays.

\subsection{WIYN RV Measurements \& Precision} \label{subsec:wiyn_precision}
RVs for single-lined stars are calculated using the \textsf{IRAF} task \textsf{fxcor} \citep{fitzpatrick_iraf_1993}, which computes the cross-correlation function (CCF) between science and an observed solar template spectrum. The velocity shift is determined by fitting a Gaussian function to the CCF peak and then corrected for the template offset and Earth's motion to obtain heliocentric velocities. To reduce erroneous measurements, we exclude observations if the CCF peak height is below 0.4. 

Stars displaying rapid rotation are fit via \textsf{fxcor} interactively. If the projected rotation velocity, $v  \sin{i}$, is greater than 120 km s$^{-1}$, we are unable to report an RV measurement due to extreme broadening of the CCF. Our spectral resolution sets a 10 km s$^{-1}$ as the lower limit on our measurement of $v \sin{i}$. 

Using the distribution of RV standard deviations of stars with small velocity variations ($\sigma_{\text{obs}} < 1.6$ km s$^{-1}$), \citetalias{gellerWIYNOpenCluster2008} find a precision measure of $\sigma_i = 0.4$ km s$^{-1}$. We adopt the same measure. 

We also maintain previous formalisms (following the analysis of \citetalias{gellerWIYNOpenCluster2008}) by classifying stars as velocity variable if their RV standard deviations (external error, $\sigma_e$ or $e$) exceed four times the precision (internal error, $\sigma_i$ or $i$), or $e/i > 4$. Specifically, for stars with low $v  \sin{i}$, those with an RV standard deviation greater than 1.6 km s$^{-1}$ are considered velocity variable. As described by \cite{geller_wiyn_2010}, for stars with $v \sin{i}$ higher than 10 km s$^{-1}$, we compute the internal error as a function of $v \sin{i}$ by: 

\begin{equation} \label{eqn:int_err}
    \sigma_i = 0.38 + (0.012 \times v \sin{i}) \text{ km s}^{-1}.
\end{equation}

\subsection{APOGEE RV Precision and Zero Point Correction} \label{subsec:zeropoint}

APOGEE RV measurements have a precision of $\sim 0.1$ km s$^{-1}$ for S/N $>$ 20 \citep{wilson_apache_2019}. We use this as the uncertainty on RV precision for APOGEE data. 

In order to check for a zero-point offset between our datasets, we use a subset of narrow-lined single stars as described in Section~\ref{subsec:wiyn_precision}. For a sample of 107 stars with at least 3 RV measurements (with an average of 10 observations each), we find an offset of $0.645 \pm 0.04$ km s$^{-1}$, where we compute the uncertainty as the standard deviation around the median. We correct for this difference by subtracting this offset value from all APOGEE measurements. 

\subsection{RV Measurement Completeness} \label{subsec: rv_completeness}

\begin{figure}[h!]
    \centering
    \includegraphics[width=\linewidth]{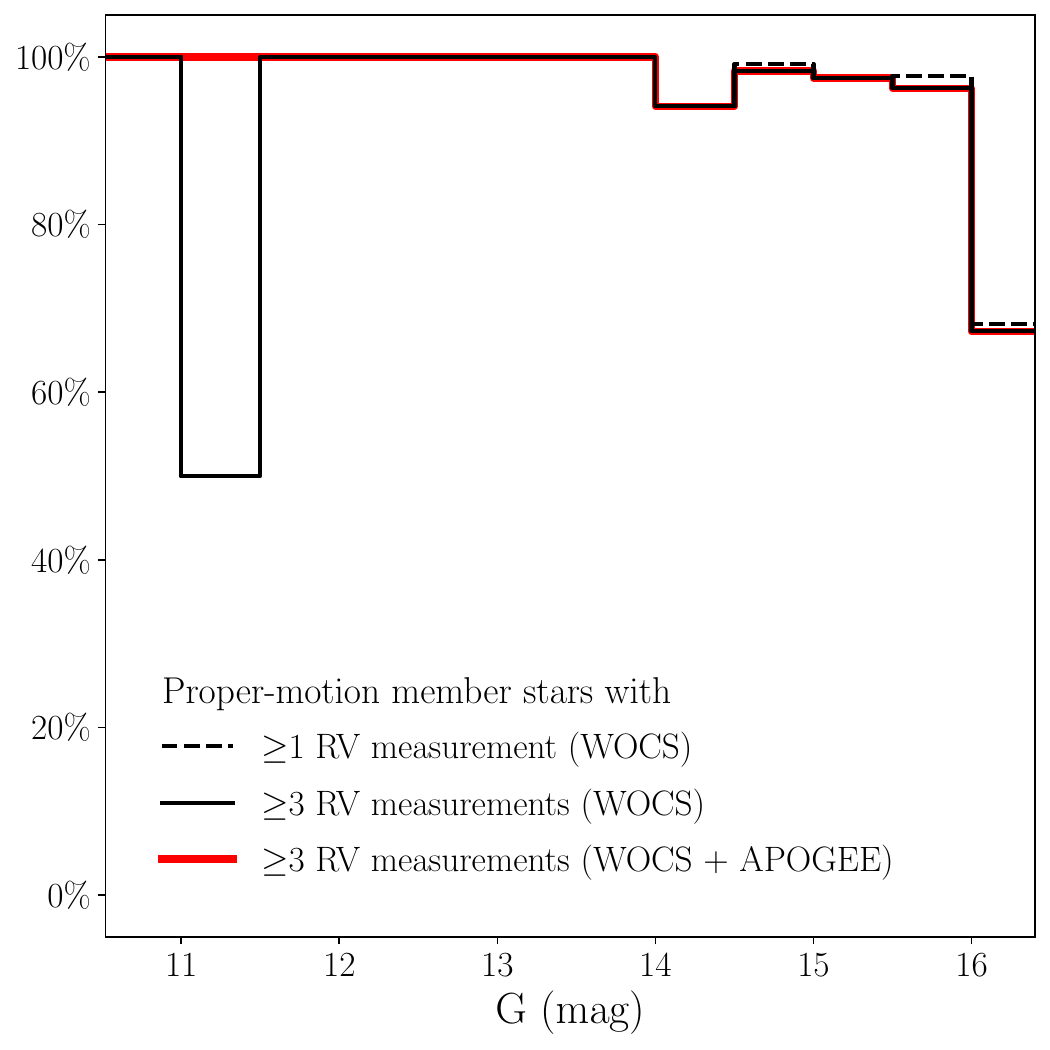}
    \caption{The percentage by G magnitude of proper-motion members that have at least one (dashed line) or three (solid line) RV measurements over a time interval of at least two years. We show the completeness of our sample with only WOCS observations (black) and our full dataset (red).}
    \label{fig:sample_mag}
\end{figure}

Figure~\ref{fig:sample_mag} shows the fraction of proper-motion members with at least one (dashed) or three (solid) RV measurements over an interval of at least two years, as a function of G with our full RV dataset. We achieve $97.6 \%$ completeness in one-epoch and $96.1\%$ completeness in three-epoch coverage from $\mathrm{10.5 \leq G < 16}$. Beyond $\mathrm{G = 16}$, the completeness curve declines due to the lower S/N at our faint limit, which motivates our choice of $\mathrm{G < 16}$ for the core binary census (Section~\ref{subsec:formalism}). Even so, down to $\mathrm{G \leq 16.5}$, we still obtain one-epoch coverage for {482} out of {528} members and three-epoch coverage for {478} out of {528} members. 

\subsection{RV Membership}\label{subsec:rv_membership} 

\begin{figure}[h!]
    \centering
    \includegraphics[width=\linewidth]{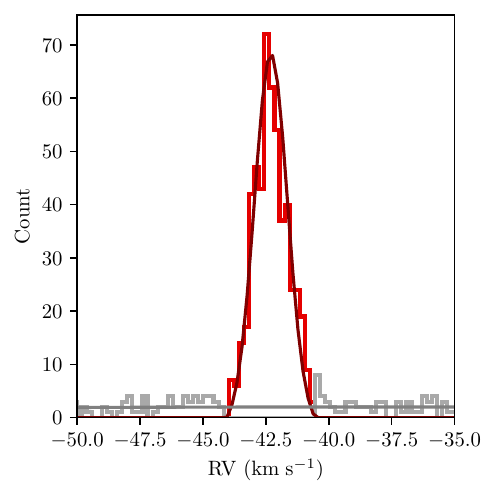}
    \caption{RV distributions of stars in the cluster (red) and the field (gray). Distributions were fit to non-velocity variable stars. Stars without orbits were binned using their mean RV measurement, meaning some member binary stars without orbit solutions may appear to not be members in this figure.}
    \label{fig:rv_gmm}
\end{figure}

We performed a GMM analysis on the 799 non-velocity-variable stars in the NGC 188 field for which we have RV measurements (see Figure~\ref{fig:rv_gmm}). The best-fit parameters for the cluster and field RV Gaussian distributions are presented in Table~\ref{table:rv_gmm_table}. 

Velocity-variable stars were given an RV membership probability based on their orbital systemic velocity if known or their mean RV measurement if the orbit solution is not known. (The latter will be conservative toward determining members.) The RV membership probability, mean RV, and RV standard deviation for each star are listed in Table~\ref{table:star_summary}.

\begin{deluxetable}{lcr}
\tablehead{\colhead{Parameter} &\colhead{\hspace{1.5cm}Cluster}
&\colhead{Field}}
\tablewidth{1.0\columnwidth}
\tabletypesize{\footnotesize}
\tablecaption{Cluster and Field Gaussian Distributions of RV\label{table:rv_gmm_table}}
\startdata
        $\mathrm{Ampl.\; (count)}$ & \hspace{1.5cm}$69.1 \pm 5.0$ & $1.89 \pm 0.14$ \\
         $\mathrm{\overline{RV}\;(km\;s^{-1})}$ & \hspace{1.5cm}$-42.29 \pm  0.04$ & $-38.06 \pm 0.05 $ \\
         $\mathrm{\sigma \;(km\;s^{-1})}$ & \hspace{1.5cm}$0.42\pm 0.08 $ & $ 1350 \pm  8$ \\
\enddata
\end{deluxetable}

\section{Cluster Members} \label{sec:membership}

\subsection{Membership Formalism} \label{subsec:formalism}

Combining the proper-motion, parallax, and RV analyses, we categorize each star in NGC 188 as members or non-members in Table~\ref{table:star_summary}. We further categorize stars as either velocity variable or not (i.e., single or binary) based on the $e/i$ value of the measurements. Following the convention of previous WOCS papers, reproduced here for convenience, we classify each star as follows:
\begin{itemize}
    \item Single Member (SM): stars that have $e/i<4$, $P_{RV} >50$\%, and $P_{\mu} >50$\%. 
    \item Single Non-member (SNM): stars that have ${e/i}<4$ and either $P_{RV} <50$\% or  $P_{\mu} <50$\%. 
    \item Binary Member (BM): velocity-variable stars (${e/i\geq4}$) that have orbital solutions, $P_{RV} >50$\%, and $P_{\mu} >50$\%. 
     \item   Binary Non-member (BNM): velocity-variable stars that have orbital solutions and either $P_{RV} <50$\% or $P_{\mu} <50$\%.
     \item   Binary Likely Member (BLM): velocity-variable stars that do not have completed orbital solutions, $P_{RV} >50$\%, and $P_{\mu} >50$\%.
     \item   Binary Likely Non-member (BLN): velocity-variable stars that do not have completed orbital solutions. Either $P_{RV} <50 \%$  or $P_{\mu} <50\%$ and the range of RVs does not include the cluster mean, making it unlikely that the orbital solution will place the star within the cluster velocity distribution.
    \item  Binary Unknown (BU): velocity-variable stars that do not have completed orbital solutions. $P_{RV} <50$\%, but $P_{\mu} >50$\%, and the range of individual RVs includes the cluster mean, making it possible that the binary could be a member. 
    \item  Very Rapid Rotator Likely Member (VRR, M): stars that are too rapidly rotating ($v\sin i > 120\mathrm{\;km\;s^{-1}}$) for accurate RV measurements, $P_{\mu} >50$\%. We include these stars as cluster members for purposes of population analyses below.
    \item  Very Rapid Rotator Likely Non-member (VRR, NM): stars that are too rapidly rotating for accurate RV measurements, $P_{\mu} <50$\%.
    \item Unknown (U): Stars with $P_{\mu} >50$\% that have fewer than three RV measurements. Some of these stars are likely RV members and/or binaries, but we do not have enough measurements to conclusively make that determination.
    \item Unknown Non-member (UNM): $P_{\mu} < 50$\% stars that have fewer than three RV measurements. 
\end{itemize}

 Table~\ref{star_category_counts} lists the number of stars in each membership category. We find {404} stars with G $<$ 16 (our completeness limit) that are members (SM, BM, BLM). As an upper bound for G $<$ 16, if we include proper-motion member binaries that have RV measurements that overlap the cluster's RV distribution (BU) and proper-motion members with fewer than 3 RV measurements, we find {419} member stars with G $<$ 16. Including stars to our observation limit of G $\leq$ 16.5 (Section~\ref{subsec: rv_completeness}), we have identified {476} member stars (SM, BM, BLM). As an upper bound for G $<$ 16.5, if we also include proper-motion member BU and U stars, we find {530} member stars. 

\begin{deluxetable}{lr}
\tablehead{\colhead{Category} &\colhead{\hspace{3cm}Count}}
\tablewidth{1.0\columnwidth}

\tablecaption{Number of Stars within each Membership Category\label{star_category_counts}}
\startdata
      SM &    346 \\
     SNM &   514 \\
      BM &     109 \\
     BNM &    60 \\
     BLM &    21 \\
     BLN &    68 \\
      BU &     5 \\
  VRR, M &      0 \\
 VRR, NM &    2 \\
       U &    49 \\
     UNM &    79 \\
\enddata
\end{deluxetable}

\subsection{Membership Changes} \label{subsec: membership_changes}

The membership classifications determined in this work are generally the same as those determined in \citetalias{geller_wiyn_2009}. However, there are a few notable membership changes between that work and this one, primarily due to improved parallax and proper-motion precisions with Gaia. 

Three previously reported binary BSSs (WOCS 451, 1888, and 7782) are not members under our analysis; the first due to being a foreground star and outside the cluster proper-motion distribution and the latter two due to having a very low probability of proper-motion membership ($\mathrm{P}_\mu< 10^{-6}$). One previously reported member single-star BSS, WOCS 1366, is not a proper-motion member. WOCS 4945---the faintest BSS of the cluster, WOCS 4540---a BSS with a secure WD companion modeled by \cite{sun_wocs_2023}, and WOCS 5020---a very long period eccentric binary---have orbital gamma velocities of $-40.5 \pm 0.7$ km $\mathrm{s^{-1}}$, $-40.4 \pm 0.3$ km $\mathrm{s^{-1}}$, and $-44.5 \pm 1.0$ km $\mathrm{s^{-1}}$ respectively, and RV membership probabilities of 46\%, 33\%, and 13\%, respectively, putting them outside the bounds of our membership criteria, yet overlapping the cluster's velocity at $<1\sigma$. These stars are proper-motion members (96\%, 92\%, and 97\%, respectively), within $3\sigma$ errors on the cluster parallax distribution ($1894.6_{-130.9}^{+151.9}$ pc,$1941.2_{-41.8}^{+43.72}$ pc, and $1960.8^{39.2}_{-37.7}$, respectively), and located in projection near the center of the cluster ($0.04^\circ$, $0.08^\circ$, and $0.02^\circ$ from the cluster center, respectively). We suspect that all three are in fact binary cluster members and list them as such, but note these should be considered BSS candidates. See Section~\ref{subsec:wocs5020} for additional discussion on the RV membership of WOCS 5020. WOCS 4447 is a binary in the BSS region with an unknown cluster membership. The RV range of the star overlaps that of the cluster, it has a 99\% proper-motion membership probability, and it is within the cluster in three dimensions spatially. This star should also be considered a BSS candidate until an orbital solution is found.

Seven additional previous member binary stars with orbit solutions do not satisfy our membership criteria. Five of the stars (WOCS 4390, 5700, 5356, 4595, and 4524) have high RV membership probabilities ($\geq 95 \%$) but are at the boundary of the cluster proper-motion distribution (Figure~\ref{fig:pmm_analysis}), so may be associated with the cluster. \cite{hunt_improving_2023} used HDBSCAN to determine cluster membership of NGC 188 using Gaia DR3 positions, parallax, and proper motions. They determined a membership probability between 33\% and 42\% for these five stars. WOCS 5762 has an orbital gamma velocity of $-40.5$ $\mathrm{km\;s^{-1}}$, which gives it an RV membership probability of 41\%, formally outside the bounds of our membership criteria, although with high proper-motion membership probability (99\%) and spatially within the cluster in three dimensions. Like with WOCS 4945 and 4540, we suspect that WOCS 5762 is also a binary cluster member and list it as such. WOCS 5242 is identified as a background star from its parallax. 

\subsection{Isochrone Fit}\label{subsec:isochrone}

\begin{figure}[h!]
    \centering
    \includegraphics[width=\linewidth]{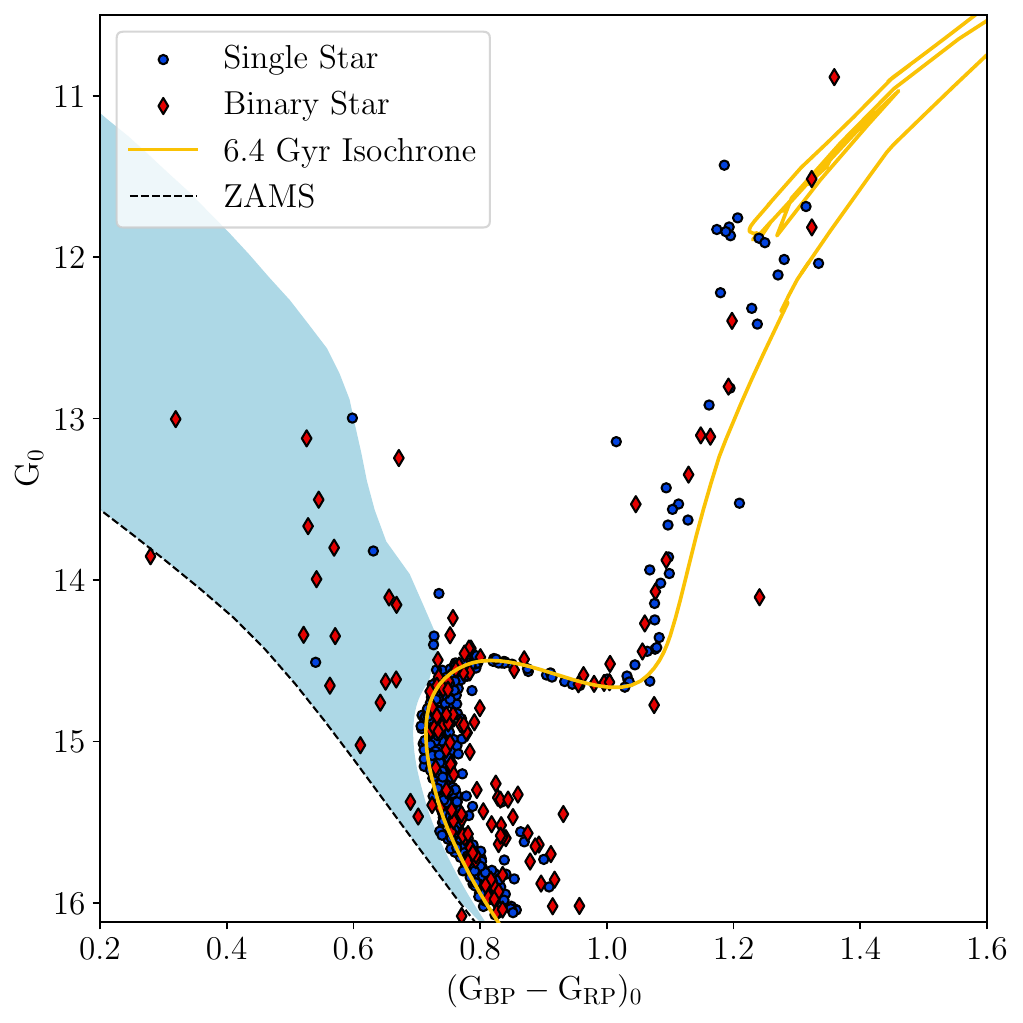}
    \caption{The color-magnitude diagram of member stars of NGC 188. Single stars are plotted as blue circles and binary stars are plotted as red diamonds. Stars have been differentially dereddened using the SFD dustmap along with an additional correction of $\mathrm{ E(G_{BP}-G_{RP})} = 0.0089$ from our isochrone fit. The zero-age main sequence is plotted as a dashed line. Our best fit \textsf{MIST} isochrone of 6.4 Gyr, [Fe/H] = $-0.024$ is shown as a solid line. The BSS region of NGC 188 is shaded in light blue. The upper bound is the terminal-age main sequence.}
\label{fig:isochrone}
\end{figure}

To determine the age and characteristics of stars in NGC 188, member stars were first differentially dereddened using the \textsf{dustmaps} Python package \citep{green_dustmaps_2018} and the Schlegel, Finkbeiner, and Davis (SFD) dustmap (\citeyear{schlegel_maps_1998}, recalibrated by \cite{schlafly_measuring_2011}). From SFD, the cluster has a mean $\mathrm{E(G_{BP}-G_{RP}) = 0.122}$ with a dispersion of 0.011 and range of 0.084 mag. We then selected non-photometric-binary members on the main sequence, subgiant branch, red giant branch, and red clump by inspection. As seen in Figure~\ref{fig:isochrone}, NGC 188 has a narrow, well-defined single-star main sequence and giant branch; we removed any star to the red of the dense population of stars along this well-defined ridge.

We fit the selected stars with \textsf{MIST} isochrones \citep{dotter_mesa_2016, choi_mesa_2016, paxton_modules_2011, paxton_modules_2013,paxton_modules_2015} having ages between 5.5 and 7 Gyr and [Fe/H] between $-0.05$ and 0.18, based on values for NGC 188 found in the literature \citep{hobbs_metallicity_1990,hippel_wiyn_1998,meibom_age_2009,casamiquela_abundanceage_2021, sun_wiyn_2022}. We use the \textsf{isochrones} Python package \citep{morton_isochrones_2015} to interpolate isochrones between ages and metallicities in the \textsf{MIST} isochrone grid and calculate apparent magnitudes and colors for distances between 1790 and 1890 pc and a small universal reddening correction. For each isochrone, we calculated the distance in color-magnitude space between every star and the nearest point on the isochrone. Our metric of best fit was the root mean square error (RMSE), with single stars on the upper main sequence and subgiant branch weighted higher than those on the giant branch as the luminosity of the subgiant branch and the shape and location of the turn-off are highly sensitive to the age of the cluster. 

We used a genetic algorithm to efficiently search the parameter space. After an initial grid search of 50,000 isochrones across our parameter space, we ran 200,000 isochrone fits, stopping once the RMSE distribution of subsequent generations converged. Each generation was created by sampling the top 15\% of the previous generation's best fit isochrones and making small perturbations to the parameters (mutation, rather than crossover). Hyperparameters for this algorithm (e.g., initial grid resolution, number of generations, the number of parents allowed to have offspring, and the size of mutations) were selected by trial and error for computational efficiency that fully explored the parameter at a resolution that potentially could be measured (e.g., 1 pc in distance, 0.005 in [Fe/H]). Our reported parameters are the best fit across all generations (i.e., the best fit parameters survived between generations), which theoretically means our population should converge toward the optimal solution \citep{rudolph_convergence_1994,eiben_global_1991}. 

Genetic algorithms are known to sample the parameter space where the probability density is high, but ultimately the sample distribution is unknown, so they may not fully represent a Bayesian posterior \citep{sambridge_geophysical_1999}, meaning that the standard deviations on the parameter distribution from our converged generations may not capture the low-probability portions of the error distribution. As such, we report ranges on each parameter (age, [Fe/H], distance, and $\mathrm{E(G_{BP} - G_{RP})}$) that were estimated from the parameter distribution of the converged generations. 

We find NGC 188 to be $6.4 \pm 0.2$ Gyr old, which falls within the age estimates found for NGC 188 in the literature (5.78 Gyr; \citealt[][]{hills_bayesian_2015},  6.17 Gyr; \citealt[][]{childs_goodbye_2024}, 6.2 Gyr; \citealt[][]{meibom_age_2009}, 6.41 Gyr; \citealt[][]{yakut_novel_2025}, 7.2 Gyr; \citealt[]{friel_abundances_2010}, and 7.65 Gyr; \citealt[][]{deniz_sed_2024}).

The distance we determined, $1820 \pm 5$ pc, aligns well with the Gaia-based measurement by \citep{bailer-jones_estimating_2021} of $1815 \pm 79$ pc (geometric distance for member stars brighter than G = 14) and is also consistent within uncertainties with the eclipsing binary-derived distance of $1770 \pm 75$ pc \citep{meibom_age_2009}. 

Our best fit isochrone is plotted with cluster members in Figure~\ref{fig:isochrone}. Our \textsf{MIST} isochrone fit successfully reproduces the main sequence and subgiant branch regions of NGC 188. However, the brightest giants and the red clump exhibit color offsets, similar to those reported by \cite{reyes_isochrone_2024} and \citet{linck_wiyn_2024} for other clusters. We investigated fitting the isochrone without including the red giant branch and found that it did not impact our fit. These discrepancies are indicative of known challenges in stellar evolutionary models, particularly regarding convective efficiency, opacity treatments, and core helium-burning phases \citep{reyes_isochrone_2024}.

\subsection{Blue Straggler Stars}

BSSs are defined as stars that have gained mass during their main sequence evolution and continue to burn hydrogen in their cores \citep{mathieu_blue_2025}. While they were historically identified as stars that were bluer and brighter than the main sequence turnoff (MSTO) of an associated coeval population, there also exist stars that are bluer, yet not brighter than the associated MSTO. Following \citet{mathieu_blue_2025}, we define the BSS region as the domain between the zero-age main sequence (ZAMS) and the terminal-age main sequence or the isochrone, here determined from the \textsf{MIST} tracks using the cluster parameters derived from our isochrone fit. We find that NGC 188 has 18 secure BSSs: WOCS 2679, 4230, 4290, 4306, 4348, 4535, 4581, 4589, 4970,  5078, 5325, 5350, 5379, 5434, 5467, 5885, 5934, and 8104. Four other stars in the BSS region of the color-magnitude diagram (CMD), WOCS 4447, 4540, 4945 and 5020, may be member stars (see Section~\ref{subsec: membership_changes}) and thus are candidate BSS. NGC 188 also has one luminous (G $\approx 13$ mag) binary just to the red of the BSS region (WOCS 5671), making this a yellow straggler star. All of these 23 stars are plotted in Figure~\ref{fig:isochrone} in (or next to) the BSS region. Of these 23 stars, 20 are binary stars, giving a BSS binary fraction of $87 \pm 19\%$.

The 4 stars immediately above the main sequence turn-off (WOCS 5489, 5101, 5080, 5269) are excluded as BSS candidates as they are within 0.75 mag of the turn-off magnitude and could be photometric binary stars (although 3 of the 4 would be at periods longer than we can detect). WOCS 5080 is a double-lined binary, explaining its separation from the cluster isochrone.

Finally, we note the presence of the sdB proper-motion member WOCS 4918. Our 10 observations of this star yielded no RV measurements.

\subsection{Main-Sequence Binary Frequency} \label{subsec:binary_fraction}

From the main-sequence turnoff (G = 14.3) through our completeness limit (G = 16) we identify {306} member stars (SM, BM, and BLM) in the cluster, {75} of which we also identify as spectroscopic binaries (BM and BLM), having periods $< 5000$ days. This yields an observed main-sequence binary frequency of {$24.5 \% \pm 2.8 \%$}, no different than the prior frequency measurement of $23 \% \pm 2\%$ \citepalias{geller_wiyn_2012} in our previous sample of NGC 188 member stars. The slight differences can be accounted for by improvements in membership from Gaia astrometry, new orbital solutions, and three new main-sequence binary members detected with additional RV data.

Given the WOCS RV measurement precision, detection of binaries (and orbital solutions) decreases as binary period approaches $10^4$ days (see Figure~1 in \citetalias{geller_wiyn_2012}), with a negligible fraction of binaries with periods greater than $10^4$ days is detectable. Here we correct for incompleteness in detected binaries to obtain the true binary frequency for  $P < 10^4$ days. 

Given that this RV survey of NGC 188 has a similar duration, number of observations, and sample size as the WOCS survey of M67 \citep{geller_stellar_2021}, we adopt the same binary detection completeness corrections. Through a Monte Carlo simulation of the RVs from the field-binary distribution sampled at the cadence of our observation campaign, they find a binary completeness fraction of 74\% for periods less than $10^4$ days. This gives us an incompleteness-corrected binary frequency of $33.1\% \pm 3.8 \%$. Again, this fraction is within the errors of the previously determined true binary fraction for NGC 188 of $32\% \pm 4\%$ (\citetalias{geller_wiyn_2009}, \citealt{geller_stellar_2021}), similarly corrected for periods less than $10^4$ days. 

\section{Spectroscopic Binary Orbits} \label{sec:orbits}

\subsection{Methodology} \label{subsec:methods}
To seek orbital solutions, we use \textsf{The Joker}, a Monte Carlo rejection sampler that specializes in solving binary orbits with sparse, uneven sampling \citep{price-whelan_joker_2017}. While a powerful tool for sparse data, there remain orbital degeneracies with fewer data. Recent modeling work by \cite{pan_modeling_2024} found that, independent of phase coverage, \textsf{The Joker} requires 8-10 RV data to assure a solution that converges in all parameters. 

We created a pipeline to apply \textsf{The Joker} to our sample of 483 stars with more than 8 RV measurements in the NGC 188 field. For each star, we search a parameter space with 100 million prior samples in a period range between 1 day and 20,000 days (the maximum period for which WOCS is sensitive to detecting RV variability for an equal mass, edge-on circular binary of masses relevant to NGC 188).

Using the rejection samples as priors, we then run a Markov chain Monte Carlo (MCMC) sampler to further explore how well different combinations of orbital parameters reproduce our RV data and guard against settling in a local minimum. Finally, we compute our final orbit solution using a direct-integrator that uses numerical techniques to minimize residuals \citep{conti_are_1973,  latham_survey_2002, goldberg_survey_2002}. 

\subsection{Single-lined Orbital Solutions} \label{subsec:orbit_sols}
In Table~\ref{table:orbits}, we provide orbital parameters for 35 member binaries with newly identified orbits. For each star, the first row includes the WOCS ID, the orbital period ($P$), the number of cycles observed, the center-of-mass RV ($\gamma$), the semi-amplitude velocity (K), the eccentricity ($ecc$), the longitude of periastron ($\omega$), a reduced Julian date of periastron passage (T$_0$), the projected semi-major axis ($a \sin{(i)}$), the mass function ($f(m)$), the root-mean-square residual velocity from the orbital solution ($\sigma$), and the number of RV measurements (N). Where applicable, the errors associated with these values are presented in the row below. We show the orbital solutions for each binary in Figure~\ref{fig:sb1fig1}, plotting the orbit curve in the top panel and the RV residuals in the bottom panel. 

Similarly, in the \customref{Appendix}{sec:appendix}, in Table~\ref{table:field_orbits} we present the orbital parameters of 44 SB1 field binaries serendipitously discovered in the NGC 188 field, and show there the orbital solutions for each field binary in Figure~\ref{fig:fsb1fig1}. 

\startlongtable
\begin{deluxetable*}{cccccccccccc} \label{table:orbits}
\tablewidth{0pt}
\tabletypesize{\scriptsize}
\tablecaption{Orbital Parameters of NGC 188 Single-Lined Binaries}

\tablehead{
   \colhead{ID} & \colhead{P} & \colhead{Orbital} & \colhead{$\gamma$} & \colhead{K} & \colhead{$ecc$} & \colhead{$\omega$} & \colhead{T$_0$} & \colhead{$a \; \sin(i)$} & \colhead{$f(m)$} & \colhead{$\sigma$} & \colhead{N} \\
   \colhead{} & \colhead{(days)} & \colhead{Cycles} & \colhead{(km s$^{-1}$)} & \colhead{(km s$^{-1}$)} & \colhead{} & \colhead{(deg)} & \colhead{(HJD-2,400,000 d)} & \colhead{($10^6$ km)} & \colhead{($M_\odot$)} & \colhead{(km s$^{-1}$)} & \colhead{}
}

\startdata
269 & 553 & 11.3 & -42.1 & 7.1 & 0.72 & 69 & 55313 & 38 & $7.0*10^{-3}$ & 0.886 & 15 \\ 
 & $\pm$ 3 &  & $\pm$ 0.4 & $\pm$ 0.9 & $\pm$ 0.05 & $\pm$ 12 & $\pm$ 17 & $\pm$ 6 & $\pm$ $3.0*10^{-3}$ &  &  \\ 
583 & 119.78 & 55.1 & -41.22 & 6.27 & 0.18 & 64 & 55474 & 10.2 & $2.9*10^{-3}$ & 0.618 & 20 \\ 
 & $\pm$ 0.04 &  & $\pm$ 0.14 & $\pm$ 0.22 & $\pm$ 0.04 & $\pm$ 12 & $\pm$ 4 & $\pm$ 0.4 & $\pm$ $3.0*10^{-4}$ &  &  \\ 
3719 & 51.5745 & 128.6 & -43.06 & 28.1 & 0.562 & 225.8 & 56022.17 & 16.46 & $6.7*10^{-2}$ & 0.959 & 23 \\ 
 & $\pm$ 0.0016 &  & $\pm$ 0.22 & $\pm$ 0.4 & $\pm$ 0.008 & $\pm$ 1.4 & $\pm$ 0.12 & $\pm$ 0.24 & $\pm$ $3.0*10^{-3}$ &  &  \\ 
3755 & 10.32601 & 642.2 & -42.6 & 39.1 & 0.206 & 72 & 55602.72 & 5.43 & $5.98*10^{-2}$ & 1.083 & 19 \\ 
 & $\pm$ 0.00009 &  & $\pm$ 0.3 & $\pm$ 0.4 & $\pm$ 0.011 & $\pm$ 3 & $\pm$ 0.08 & $\pm$ 0.06 & $\pm$ $1.8*10^{-3}$ &  &  \\ 
3953 & 940.4 & 7.1 & -41.14 & 4.51 & 0.04 & 240 & 56340 & 58.2 & $8.9*10^{-3}$ & 0.633 & 24 \\ 
 & $\pm$ 2.5 &  & $\pm$ 0.14 & $\pm$ 0.19 & $\pm$ 0.04 & $\pm$ 70 & $\pm$ 170 & $\pm$ 2.5 & $\pm$ $1.1*10^{-3}$ &  &  \\ 
4022 & 54.8616 & 104.7 & -42.79 & 23.9 & 0.471 & 233 & 55957.7 & 15.9 & $5.4*10^{-2}$ & 0.714 & 15 \\ 
 & $\pm$ 0.0021 &  & $\pm$ 0.23 & $\pm$ 0.4 & $\pm$ 0.016 & $\pm$ 3 & $\pm$ 0.4 & $\pm$ 0.3 & $\pm$ $3.0*10^{-3}$ &  &  \\ 
4038 & 1361 & 4.9 & -42.83 & 9.4 & 0.58 & 271 & 56435 & 143 & $6.3*10^{-2}$ & 0.544 & 22 \\ 
 & $\pm$ 3 &  & $\pm$ 0.21 & $\pm$ 0.5 & $\pm$ 0.05 & $\pm$ 3 & $\pm$ 7 & $\pm$ 10. & $\pm$ $1.1*10^{-2}$ &  &  \\ 
4122 & 41.3357 & 138.5 & -42.4 & 27.9 & 0.423 & 206 & 55777.4 & 14.4 & $6.9*10^{-2}$ & 1.012 & 15 \\ 
 & $\pm$ 0.0014 &  & $\pm$ 0.3 & $\pm$ 0.7 & $\pm$ 0.014 & $\pm$ 3 & $\pm$ 0.3 & $\pm$ 0.4 & $\pm$ $5.0*10^{-3}$ &  &  \\ 
4125 & 4900 & 1.8 & -42.72 & 6.04 & 0.439 & 161 & 50890 & 366 & $8.1*10^{-2}$ & 0.518 & 41 \\ 
 & $\pm$ 40 &  & $\pm$ 0.08 & $\pm$ 0.15 & $\pm$ 0.022 & $\pm$ 3 & $\pm$ 40 & $\pm$ 10. & $\pm$ $6.0*10^{-3}$ &  &  \\ 
4187 & 7.94637 & 834.5 & -41.2 & 27.0 & 0.267 & 258 & 55928.46 & 2.84 & $1.45*10^{-2}$ & 1.054 & 18 \\ 
 & $\pm$ 0.00006 &  & $\pm$ 0.3 & $\pm$ 0.4 & $\pm$ 0.017 & $\pm$ 4 & $\pm$ 0.08 & $\pm$ 0.04 & $\pm$ $7.0*10^{-4}$ &  &  \\ 
4230 & 0.455609 & 21391.0 & -40.8 & 43.6 & 0.04 & 240 & 53913.97 & 0.273 & $3.9*10^{-3}$ & 4.256 & 32 \\ 
 & $\pm$ 0.0 &  & $\pm$ 0.8 & $\pm$ 1.0 & $\pm$ 0.03 & $\pm$ 40 & $\pm$ 0.05 & $\pm$ 0.006 & $\pm$ $3.0*10^{-4}$ &  &  \\ 
4508 & 2.043215 & 3182.1 & -42.00 & 20.1 & 0.033 & 344 & 54774.64 & 0.564 & $1.72*10^{-3}$ & 0.744 & 17 \\ 
 & $\pm$ 0.000009 &  & $\pm$ 0.22 & $\pm$ 0.3 & $\pm$ 0.018 & $\pm$ 22 & $\pm$ 0.12 & $\pm$ 0.009 & $\pm$ $8.0*10^{-5}$ &  &  \\ 
4523 & 97.937 & 58.5 & -42.17 & 23.8 & 0.696 & 147.1 & 55945.09 & 23.0 & $5.0*10^{-2}$ & 0.823 & 22 \\ 
 & $\pm$ 0.005 &  & $\pm$ 0.18 & $\pm$ 0.4 & $\pm$ 0.008 & $\pm$ 1.6 & $\pm$ 0.24 & $\pm$ 0.5 & $\pm$ $3.0*10^{-3}$ &  &  \\ 
4593 & 64.829 & 100.3 & -43.4 & 20.3 & 0.25 & 223 & 55666 & 17.6 & $5.1*10^{-2}$ & 4.325 & 22 \\ 
 & $\pm$ 0.024 &  & $\pm$ 1 & $\pm$ 1.7 & $\pm$ 0.08 & $\pm$ 18 & $\pm$ 3 & $\pm$ 1.5 & $\pm$ $1.3*10^{-2}$ &  &  \\ 
4609 & 559.5 & 15.6 & -42.09 & 6.0 & 0.45 & 61 & 54414 & 41 & $9.0*10^{-3}$ & 0.937 & 23 \\ 
 & $\pm$ 1 &  & $\pm$ 0.21 & $\pm$ 0.6 & $\pm$ 0.07 & $\pm$ 9 & $\pm$ 9 & $\pm$ 4 & $\pm$ $3.0*10^{-3}$ &  &  \\ 
4627 & 1507.3 & 4.3 & -41.5 & 12.4 & 0.812 & 241 & 56263 & 150. & $5.9*10^{-2}$ & 1.323 & 16 \\ 
 & $\pm$ 1.0 &  & $\pm$ 0.4 & $\pm$ 0.6 & $\pm$ 0.017 & $\pm$ 6 & $\pm$ 3 & $\pm$ 10. & $\pm$ $1.0*10^{-2}$ &  &  \\ 
4843 & 1241.7 & 6.3 & -42.29 & 5.61 & 0.163 & 253 & 56741 & 94.5 & $2.18*10^{-2}$ & 0.314 & 54 \\ 
 & $\pm$ 1.1 &  & $\pm$ 0.05 & $\pm$ 0.06 & $\pm$ 0.010 & $\pm$ 5 & $\pm$ 15 & $\pm$ 1 & $\pm$ $7.0*10^{-4}$ &  &  \\ 
4858 & 605.9 & 10.9 & -40.86 & 9.8 & 0.54 & 338 & 56050. & 69 & $3.5*10^{-2}$ & 0.916 & 24 \\ 
 & $\pm$ 0.9 &  & $\pm$ 0.20 & $\pm$ 0.5 & $\pm$ 0.03 & $\pm$ 4 & $\pm$ 6 & $\pm$ 4 & $\pm$ $5.0*10^{-3}$ &  &  \\ 
4945 & 5110 & 1.1 & -40.5 & 7.3 & 0.48 & 253 & 54330 & 450 & $1.4*10^{-1}$ & 0.917 & 15 \\ 
 & $\pm$ 50 &  & $\pm$ 0.7 & $\pm$ 1.2 & $\pm$ 0.07 & $\pm$ 9 & $\pm$ 90 & $\pm$ 70 & $\pm$ $7.0*10^{-2}$ &  &  \\ 
5020 & 38000 & 0.3 & -44.5 & 3.3 & 0.74 & 329 & 51850 & 1100 & $4.0*10^{-2}$ & 0.347 & 27 \\ 
 & $\pm$ 7000 &  & $\pm$ 1 & $\pm$ 0.3 & $\pm$ 0.16 & $\pm$ 9 & $\pm$ 150 & $\pm$ 300 & $\pm$ $5.0*10^{-2}$ &  &  \\ 
5084 & 58.645 & 116.6 & -42.9 & 27.7 & 0.497 & 185.0 & 53998.4 & 19.4 & $8.4*10^{-2}$ & 0.513 & 17 \\ 
 & $\pm$ 0.006 &  & $\pm$ 0.3 & $\pm$ 0.5 & $\pm$ 0.012 & $\pm$ 1.1 & $\pm$ 0.3 & $\pm$ 0.4 & $\pm$ $5.0*10^{-3}$ &  &  \\ 
5103 & 13.4606 & 272.9 & -42.2 & 41.2 & 0.037 & 152 & 54692.8 & 7.63 & $9.8*10^{-2}$ & 0.881 & 12 \\ 
 & $\pm$ 0.0008 &  & $\pm$ 0.4 & $\pm$ 0.7 & $\pm$ 0.018 & $\pm$ 15 & $\pm$ 0.6 & $\pm$ 0.13 & $\pm$ $5.0*10^{-3}$ &  &  \\ 
5107 & 64.18 & 83.4 & -41.3 & 10.0 & 0.13 & 340 & 55376 & 8.8 & $6.6*10^{-3}$ & 1.656 & 12 \\ 
 & $\pm$ 0.03 &  & $\pm$ 0.6 & $\pm$ 0.8 & $\pm$ 0.08 & $\pm$ 40 & $\pm$ 7 & $\pm$ 0.7 & $\pm$ $1.6*10^{-3}$ &  &  \\ 
5115 & 7.86389 & 784.3 & -41.85 & 18.3 & 0.023 & 130 & 56077.9 & 1.98 & $5.0*10^{-3}$ & 1.024 & 18 \\ 
 & $\pm$ 0.00008 &  & $\pm$ 0.25 & $\pm$ 0.4 & $\pm$ 0.020 & $\pm$ 50 & $\pm$ 1.0 & $\pm$ 0.04 & $\pm$ $3.0*10^{-4}$ &  &  \\ 
5144 & 20.785 & 176.7 & -42.5 & 30. & 0.48 & 200.6 & 54768.90 & 7.6 & $4.0*10^{-2}$ & 0.776 & 9 \\ 
 & $\pm$ 0.006 &  & $\pm$ 0.4 & $\pm$ 3 & $\pm$ 0.06 & $\pm$ 2.0 & $\pm$ 0.11 & $\pm$ 0.8 & $\pm$ $1.3*10^{-2}$ &  &  \\ 
5396 & 237.66 & 24.4 & -41.6 & 11.5 & 0.38 & 257 & 56326 & 34.8 & $3.0*10^{-2}$ & 1.13 & 16 \\ 
 & $\pm$ 0.11 &  & $\pm$ 0.3 & $\pm$ 0.5 & $\pm$ 0.04 & $\pm$ 7 & $\pm$ 3 & $\pm$ 1.7 & $\pm$ $4.0*10^{-3}$ &  &  \\ 
5454 & 757.25 & 12.6 & -41.32 & 14.8 & 0.750 & 339.5 & 53364.1 & 101.7 & $7.3*10^{-2}$ & 0.597 & 48 \\ 
 & $\pm$ 0.14 &  & $\pm$ 0.09 & $\pm$ 0.3 & $\pm$ 0.007 & $\pm$ 1.1 & $\pm$ 0.7 & $\pm$ 2.3 & $\pm$ $5.0*10^{-3}$ &  &  \\ 
5497 & 831 & 7.8 & -41.6 & 4.0 & 0.19 & 40 & 55270 & 44 & $5.0*10^{-3}$ & 0.746 & 13 \\ 
 & $\pm$ 5 &  & $\pm$ 0.3 & $\pm$ 0.6 & $\pm$ 0.12 & $\pm$ 30 & $\pm$ 80 & $\pm$ 6 & $\pm$ $2.2*10^{-3}$ &  &  \\ 
5520 & 191.37 & 24.0 & -43.17 & 6.0 & 0.64 & 352 & 56248 & 12 & $2.0*10^{-3}$ & 0.471 & 13 \\ 
 & $\pm$ 0.06 &  & $\pm$ 0.24 & $\pm$ 1.9 & $\pm$ 0.08 & $\pm$ 7 & $\pm$ 3 & $\pm$ 4 & $\pm$ $1.8*10^{-3}$ &  &  \\ 
5627 & 107.52 & 58.1 & -42.56 & 6.6 & 0.42 & 27 & 55131.5 & 8.8 & $2.4*10^{-3}$ & 0.701 & 16 \\ 
 & $\pm$ 0.04 &  & $\pm$ 0.20 & $\pm$ 0.5 & $\pm$ 0.04 & $\pm$ 8 & $\pm$ 2.2 & $\pm$ 0.7 & $\pm$ $6.0*10^{-4}$ &  &  \\ 
5733 & 1.168511 & 2983.1 & -41.8 & 60.8 & 0.021 & 180 & 51869.56 & 0.976 & $2.72*10^{-2}$ & 3.44 & 28 \\ 
 & $\pm$ 0.000003 &  & $\pm$ 0.7 & $\pm$ 1.1 & $\pm$ 0.017 & $\pm$ 40 & $\pm$ 0.14 & $\pm$ 0.017 & $\pm$ $1.5*10^{-3}$ &  &  \\ 
6201 & 37.537 & 149.6 & -42.14 & 15.3 & 0.275 & 332 & 55963.3 & 7.60 & $1.24*10^{-2}$ & 0.823 & 18 \\ 
 & $\pm$ 0.003 &  & $\pm$ 0.24 & $\pm$ 0.4 & $\pm$ 0.020 & $\pm$ 4 & $\pm$ 0.4 & $\pm$ 0.18 & $\pm$ $9.0*10^{-4}$ &  &  \\ 
8104 & 3460 & 1.9 & -40.6 & 5.31 & 0.12 & 110 & 57900 & 251 & $5.2*10^{-2}$ & 0.763 & 25 \\ 
 & $\pm$ 100 &  & $\pm$ 0.4 & $\pm$ 0.22 & $\pm$ 0.05 & $\pm$ 30 & $\pm$ 400 & $\pm$ 10. & $\pm$ $7.0*10^{-3}$ &  &  \\ 
8406 & 2938 & 3.0 & -42.20 & 1.75 & 0.08 & 20 & 52500 & 70. & $1.61*10^{-3}$ & 0.411 & 68 \\ 
 & $\pm$ 24 &  & $\pm$ 0.06 & $\pm$ 0.07 & $\pm$ 0.04 & $\pm$ 40 & $\pm$ 300 & $\pm$ 3 & $\pm$ $2.0*10^{-4}$ &  &  \\ 
8719 & 225.44 & 25.4 & -42.0 & 7.6 & 0.28 & 40. & 56040. & 22.5 & $9.0*10^{-3}$ & 0.866 & 14 \\ 
 & $\pm$ 0.18 &  & $\pm$ 0.3 & $\pm$ 0.3 & $\pm$ 0.05 & $\pm$ 12 & $\pm$ 7 & $\pm$ 1.1 & $\pm$ $1.2*10^{-3}$ &  &  \\ 
\enddata
\end{deluxetable*}

\input{ngc188.sb1pictures}

\clearpage

Three binaries in our sample of new orbits---WOCS 5733, WOCS 4843, and WOCS 8406--- represent updates from previously published orbital solutions. WOCS 5733 now has an improved 1.16 day circular orbit derived from additional RV measurements, replacing the earlier $8.7040 \pm 0.0004$ day orbit reported by \citetalias{geller_wiyn_2009}. The revised orbits for WOCS 4843 and WOCS 8406 are consistent with earlier determinations  but offer higher precision solutions resulting from our combination of the WOCS and APOGEE datasets. 

\section{Discussion} \label{sec:discussion}

\subsection{Determination of The Tidal Circularization Period} \label{subsec:tides}

Tidal forces are an important consideration in interactions between stars in short-period orbits. The transition from eccentric to circular orbits can be explained by dissipative tidal forces that convert the orbital angular momentum to rotational momentum, tidally locking the two stars \citep{zahn_tidal_1977, hut_tidal_1981}. 

\subsubsection{Fitting the $e$---$\log(P)$ Distribution}

\cite{meibom_robust_2005} proposed a fitting function to determine the cicularization period from coevual $e$---$\log(P)$ distributions. This function was constructed to mimic the tidal circularization isochrone of the most frequently occurring eccentric binary orbits. We present this function in Equation~\ref{subeqn:tidalfit1},
\begin{subequations}
\begin{equation} \label{subeqn:tidalfit1}
e(P) = 
     \begin{cases} 
     0.0 & \text{if } P \leq P' \\ 
     \alpha (1- e^{\beta(P'-P)})^\gamma & \text{if } P > P' 
     \end{cases}
\end{equation}
\begin{equation} \label{subeqn:tidalfit2}
    \delta = \sum_{i=1}^N |e_i - e(P_i)|^\eta
\end{equation}
\end{subequations}

\noindent where $\alpha = 0.35$ is set to ensure that $e(P>P')$ approaches the mean eccentricity of all observed binary orbits with periods longer than 50 days in the Pleides, M35, Hyades, M67, and NGC 188. While $\gamma = 1.0$ controls the abruptness of the break from circularity, $\beta = 0.14$ controls the slope of the transition, with the values being adopted to minimize the width of Monte Carlo simulations. \cite{meibom_robust_2005} define the ``tidal circularization period'' as the orbital period corresponding to $e(P)=0.01$. 

Given our expanded sample of binary orbits, we revisit the circularization period of NGC 188, measured by previous studies (15.0 days; \citealt{mathieu_wiyn_2004} \& \citetalias{geller_wiyn_2012}, $14.5^{+1.4}_{-2.2}$ days \citealt{meibom_robust_2005}). 

Tidal interactions depend not only on the orbital separation between the binary companions but also sensitively on the radii of the stars and the depths of their convective zones. To limit systematic errors on the tidal circularization measurement, we use only unevolved main-sequence stars, specifically 49 stars between $\mathrm{G}=15.3$ and $\mathrm{G}=16.5$ (our observation limit). 

\cite{meibom_robust_2005} determined the circularization period by minimizing the absolute deviations of observed eccentricities from Equation \ref{subeqn:tidalfit2} with $\eta = 1$, and used Monte Carlo simulations to estimate the uncertainties. In order to account for errors in both period and eccentricity, \cite{geller_stellar_2021} developed an orthogonal distance regression (ODR) method to fit the circularization function, along with a bootstrapping technique to derive a distribution of measured circularization periods. 

However, using this technique for the binaries of NGC 188, we find degeneracies in the derived circularization periods as a function of initial guess provided to the ODR function. We instead employ an MCMC-based approach\footnote{\url{https://github.com/ritviksainarayan/tidalcircularization}} in which the likelihood is evaluated using Gauss-Hermite quadrature to determine the tidal circularization period. This approach offers several advantages: the MCMC algorithm robustly explores multimodal and degenerate likelihood surfaces where optimizers such as the ODR may become trapped, capturing non-Gaussian uncertainties in both $P$ and $e$ \citep{ford_quantifying_2005}, and the use of Gauss-Hermite quadrature enables accurate and efficient approximation of the marginal likelihood \citep{bianconcini_asymptotic_2014, stringer_inference_2025}.

\subsubsection{Circularization Results} \label{subsubsec:tidal_results}

\begin{figure}[hptb]
    \centering
    \includegraphics[width=\linewidth]{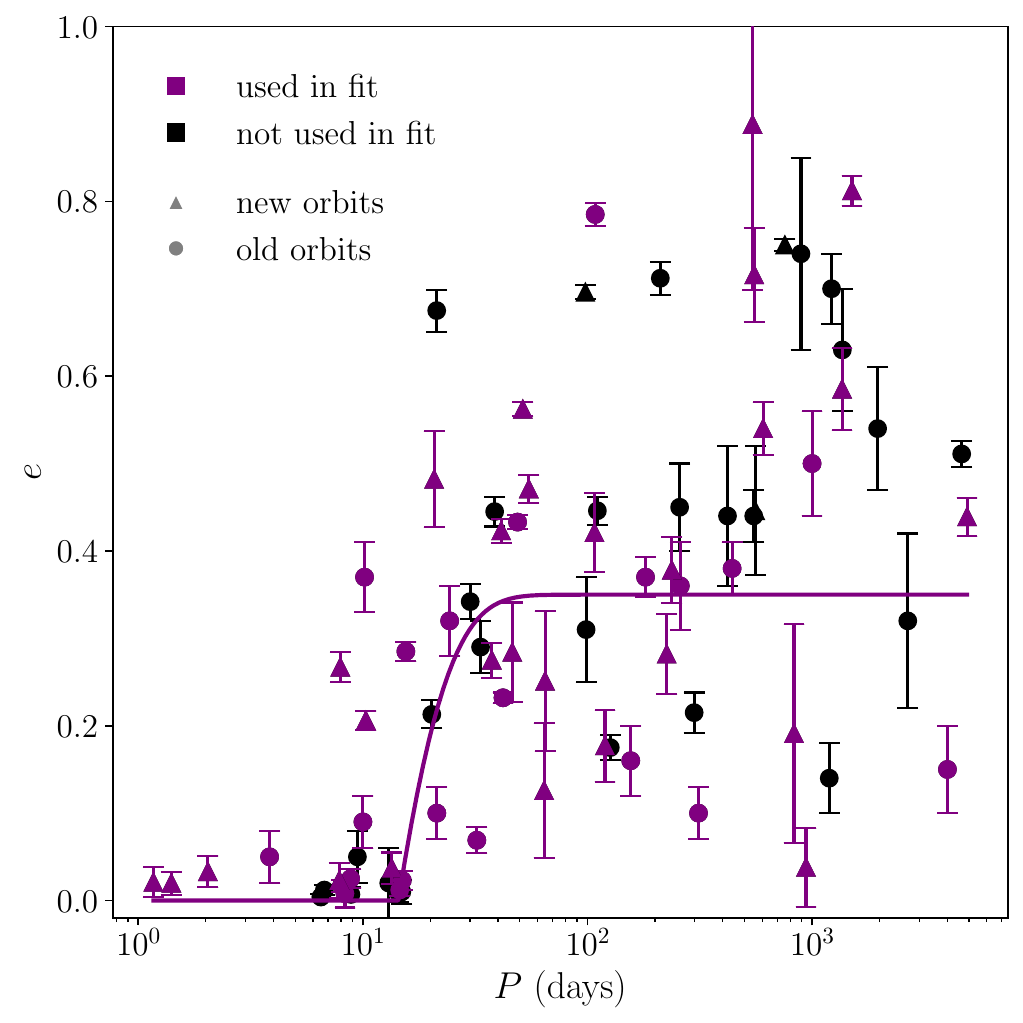}
    \caption{The period–eccentricity ($e$–$\log P$) distribution of NGC 188 main-sequence binaries. $1\sigma$ error bars are shown for all orbits. Triangles represent newly derived orbital solutions, while circles denote previously published solutions. Purple symbols indicate binaries included in the fit to the circularization period, and black symbols show systems excluded from the fit. The solid purple line marks our best-fit circularization function \citep{meibom_robust_2005}, yielding a circularization period of $P_\mathrm{circ} = 14.4^{+0.14}_{-0.11}$ days.}
    \label{fig:elogP}
\end{figure}

We show our best fit circularization period for the NGC 188 main-sequence binaries in Figure~\ref{fig:elogP}. The purple markers represent the orbits we used in our determination of the tidal circularization period, where in particular, the purple triangles represent the new binary orbits. 

\begin{figure}[hptb]
    \centering
    \includegraphics[width=\linewidth]{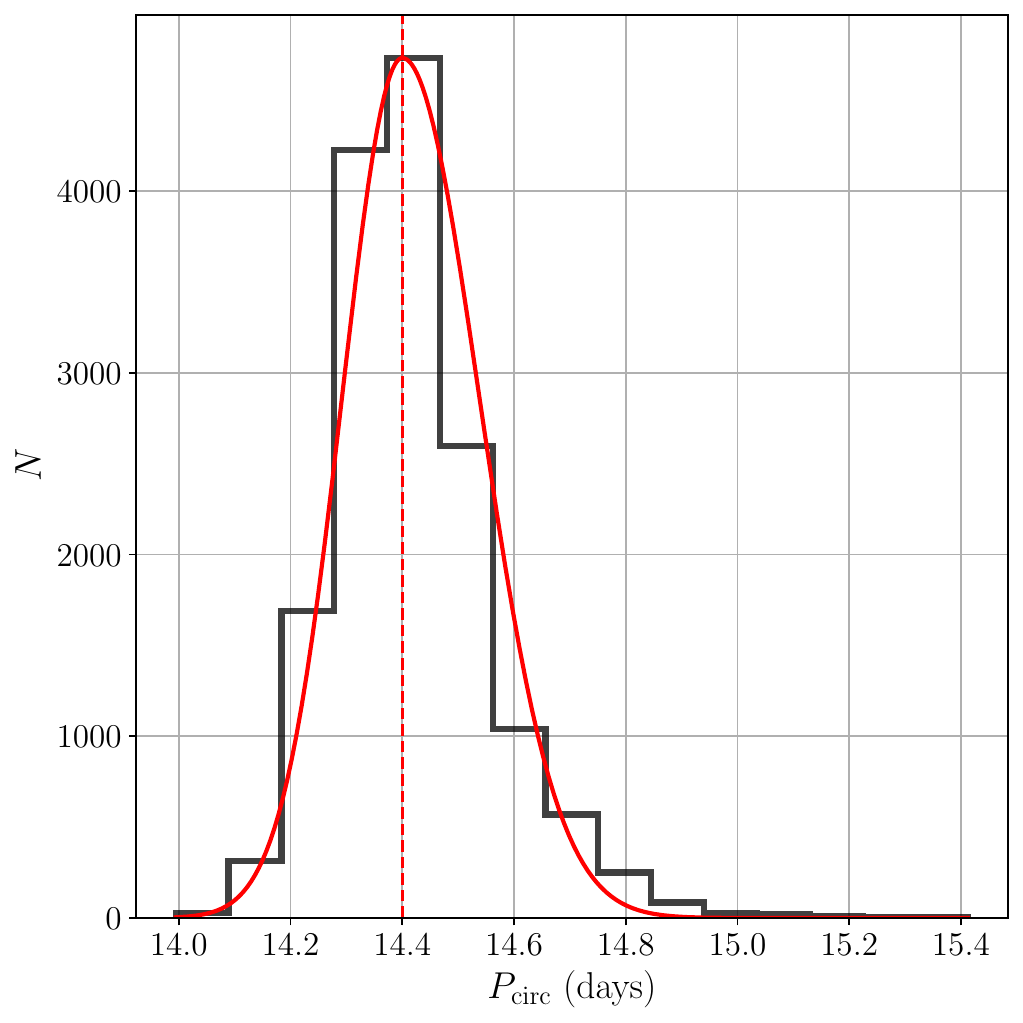}
    \caption{Distribution of circularization periods ($P_\mathrm{circ}$) resulting from the MCMC analysis, fit with an asymmetric Gaussian function. The red curve shows a $P_\mathrm{circ} = 14.4^{+0.14}_{-0.11}$ days.}
    \label{fig:pcut}
\end{figure}

The resulting distributions of circularization periods from the MCMC algorithm is shown by the black-lined histogram in Figure~\ref{fig:pcut}. Given the auto-correlation time associated with each walker, we find that using 20,000 samples with a burn-in period of 500 samples allows for model convergence with well-mixed chains. In contrast to the ODR method, the MCMC approach indicates a unimodal distribution of circularization periods at approximately 14 days. 

Following \cite{geller_stellar_2021}, we fit the distribution with an asymmetric Gaussian function, defined by a single mean and amplitude, but with two $\sigma$ values, one on either side of the mean. We measure a peak value of $P_\mathrm{circ} = 14.4^{+0.14}_{-0.11}$ days. This value is in agreement with the earlier measurements, who had a smaller sample and also included brighter binaries closer to the turnoff. 

Hence, these results are also consistent with previous work on NGC 188, which when combined with other clusters, showed that an observed increase in $P_\mathrm{circ}$ with age indicates continued tidal circularization on the main sequence, in addition to pre-main-sequence processes \citep{mathieu_wiyn_2004,meibom_robust_2005}. Recently, \citet{penev_comprehensive_2022} developed a tidal-envelope model as an alternative approach to characterize the observed period-eccentricity distributions across open clusters, including NGC 188. Using M35 as their youngest open cluster, their analysis demonstrated that all of the observed tidal circularization must occur during the pre-main-sequence phase. In contrast, for NGC 6819 and NGC 188, two older open clusters, they find that main-sequence circularization is also needed to generate the present-day period-eccentricity distribution. We report that the addition of two new main-sequence orbits with $P \approx 10$ days that have $e \sim 0.2$ and three new circular main-sequence orbits with $P < 3 $ days are both consistent with their period-eccentricity tidal-envelope model.

\subsection{The Period-Evolutionary Phase Relation} \label{subsec:period_phase}

\begin{figure}[h!]
    \centering
    \includegraphics[width=\linewidth]{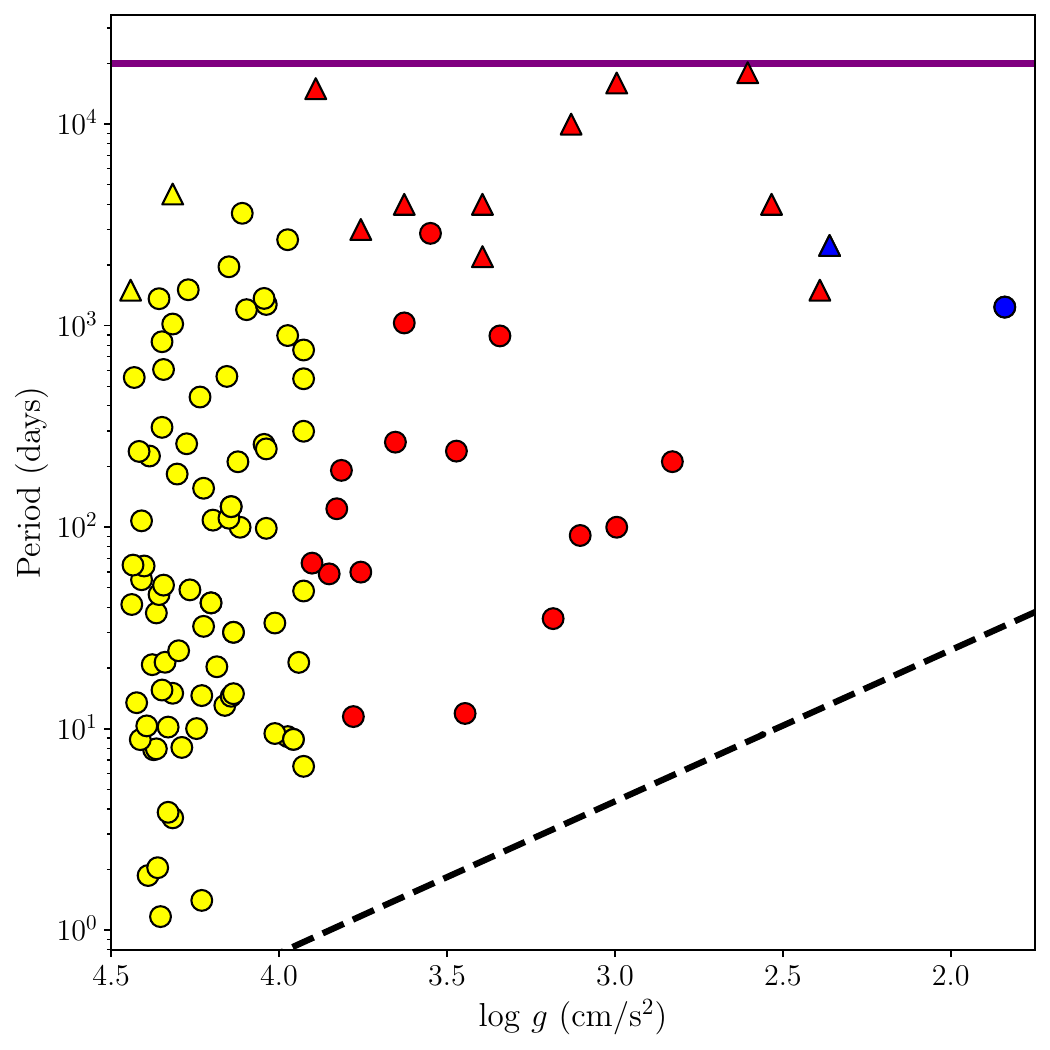}
    \caption{The distribution of orbital period against primary-star $\log{g}$ for binary stars with orbital solutions. The yellow markers represent main-sequence binaries, the red markers represent subgiant and red giant binaries, and the blue markers represent asymptotic giant binaries. Circles represent binaries that have an orbital solution whereas we use triangles to indicate a lower limit on period for the binaries that do not have an orbital solution but show long-term RV trends of at least that many days. We also show the maximum period for which WOCS is sensitive to detecting RV variability (Section~\ref{subsec:methods}) with the purple solid line. Finally, the black dashed line shows the Roche-lobe radius as a function of orbit period for equal-mass binaries. }
    \label{fig:period_phase}
\end{figure}

We use the \textsf{MIST} isochrone model for NGC 188 (Section~\ref{subsec:isochrone}) to determine theoretical stellar parameters for all cluster members (except BSSs) based on their positions on the CMD. Using $\log{g}$ as a proxy for stellar evolution phase, in Figure~\ref{fig:period_phase} we show the distribution of orbital period against $\log{g}$ for binary stars with orbital solutions. We see that the lower bound on orbital period steadily increases with decreasing $\log g$, or equivalently with a larger primary-star radius, from the main sequence through the red giant branch. Using a Kolmogorov–Smirnov test, we find that the period distribution for the main sequence and red giant populations are statistically different with a significance of $p = 0.03$.

Roche-lobe overflow (RLOF), a key mass-transfer pathway, depends sensitively on both the initial orbital separation and mass ratio, which influence not only whether RLOF occurs but also whether it proceeds stably or leads to a common-envelope phase \citep{soberman_stability_1997}. The black-dashed line in Figure~\ref{fig:period_phase} sets a lower limit on the orbital period (for an equal mass binary) as the stars evolve. We determine this lower limit by combining Kepler's third law with the Eggleton approximation \citep[Equation 2 of][]{eggleton_aproximations_1983} and stellar parameters from our NGC 188 \textsf{MIST} isochrone. It is likely that giant progenitors in even shorter-period giants interacted with their companions, and followed alternative evolutionary paths such as forming BSSs. The line approximately parallels the observed lower boundary on the orbital periods. Our observations of shorter period orbits dropping out is consistent with findings from \cite{mathieu_binary_2009} and \cite{gosnell_implications_2015} that indicated mass transfer from giant companions is a formation mechanism for long-period BSS binaries in NGC 188. 

We note that the six stars in the red clump appear to be single stars (no RV variability for $P < 20,000$ days). Along the red giant track of our \textsf{MIST} isochrone, the maximum stellar radius is $R_\mathrm{max} = 172 \; R_\odot$; for a $q=1$ circular orbit using the Eggleton approximation \citep[Equation 2 of][]{eggleton_aproximations_1983}, the Roche lobe radius corresponding to $R_{\mathrm{max}}$ implies stars with orbital periods under $P_{\rm crit} = 751$ days will interact. None of the post-red giant branch stars have periods with $P<P_{\rm crit}$, presumably due to mass transfer interactions. Of the 41 red giant members in NGC 188, 11 have or appear to have $P > P_\mathrm{ crit}$, yielding a long-period binary fraction of $27\% \pm 8\%$. The binomial theorem predicts a probability for no binary detections among the red clump stars of $15\%^{+13\%}_{-7\%}$. Of the two asymptotic giant branch stars, WOCS 4843 is a detected binary with a period of 1240 days and WOCS 6353 shows low RV variability with a period of at least 2500 days (although we report this star as a single star). Including these stars with the red clump stars in our analysis gives a long-period binary fraction that is consistent with that of the red giant branch. 

\subsection{Stars of Note}

\subsubsection{WOCS 3953: A Blue Lurker Candidate} \label{subsec:wocs3953}

\begin{figure}[h!]
    \centering
    \includegraphics[width=\linewidth]{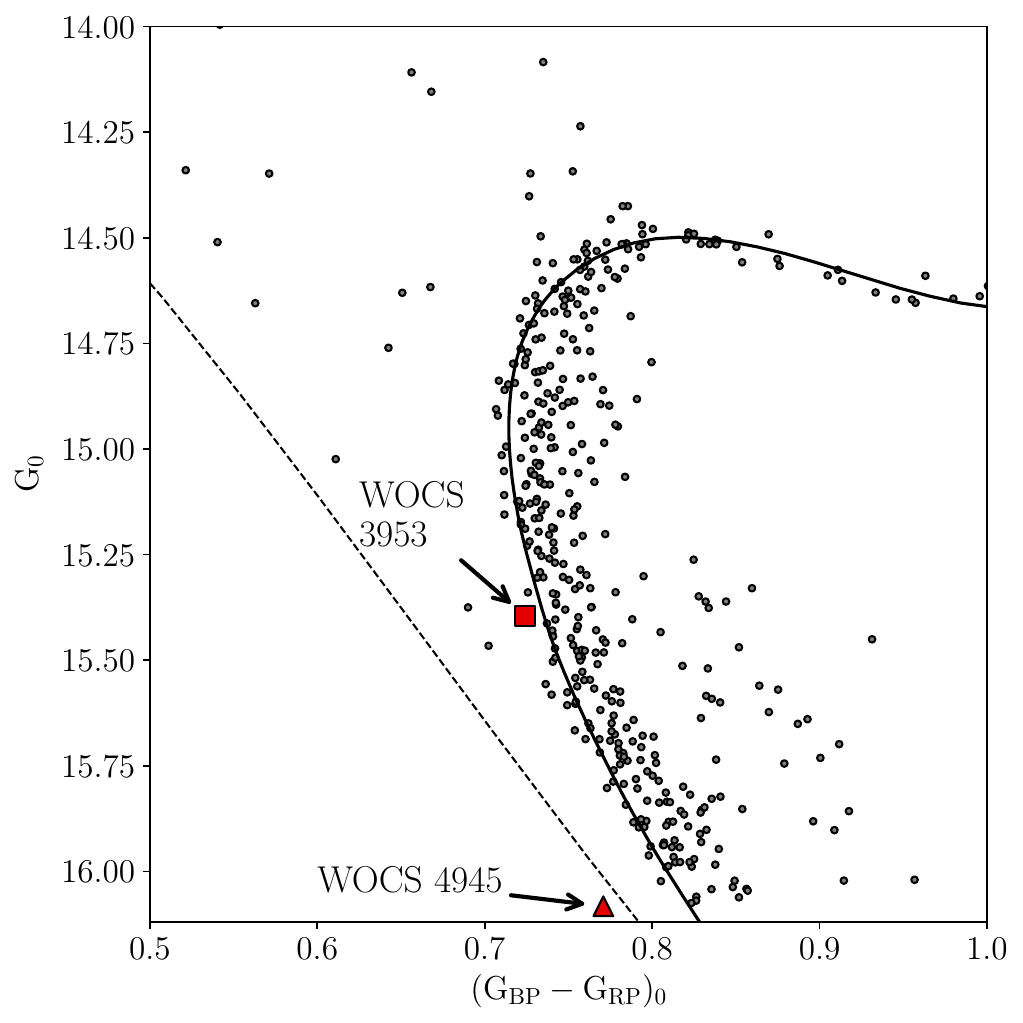}
    \caption{Position of WOCS 3953 (red square) and WOCS 4945 (red triangle) in the CMD of NGC 188.}
    \label{fig:3953_cmd}
\end{figure}

\begin{figure*}[t!]
    \centering
    \includegraphics[width=\textwidth]{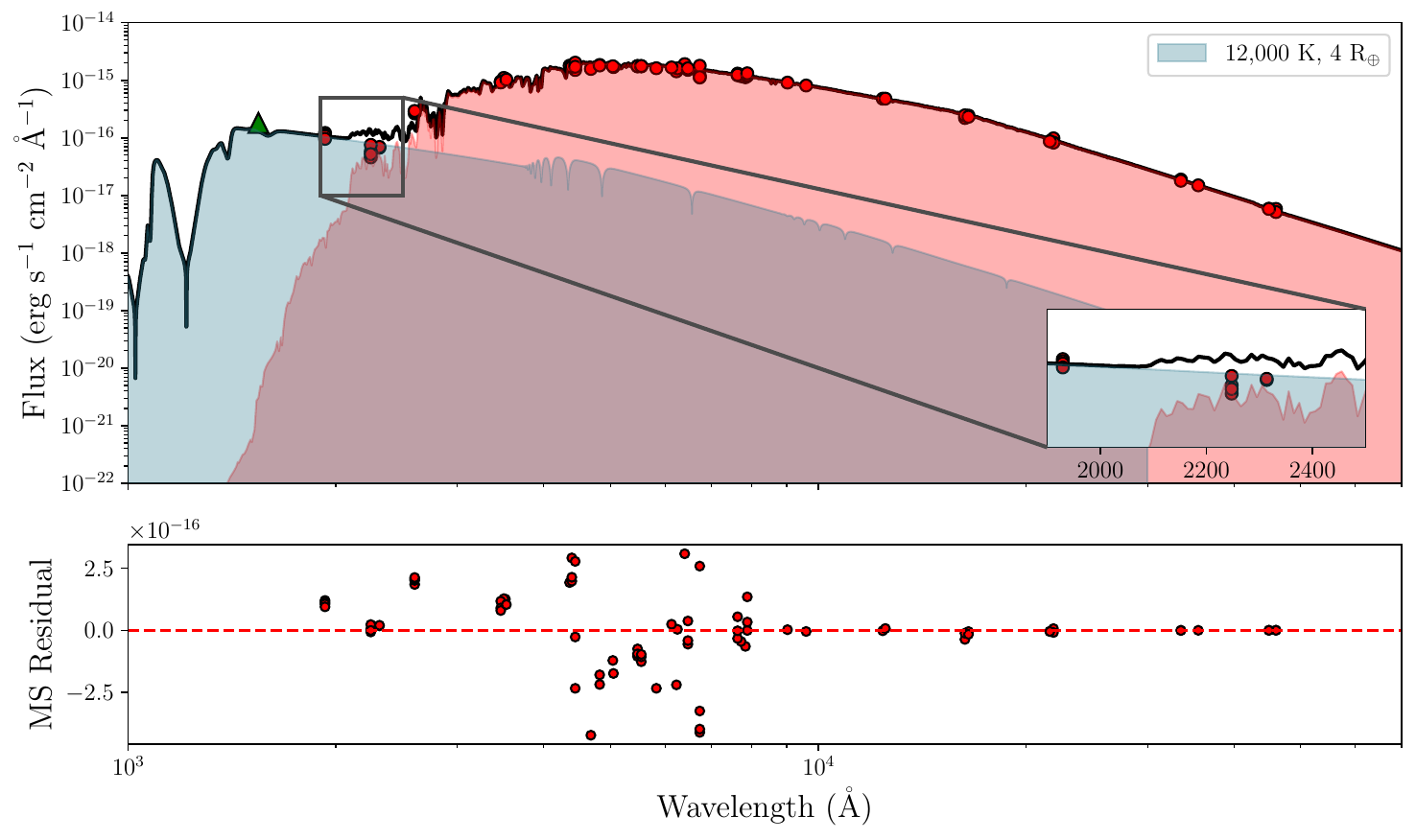}
    \caption{We show the composite SED for WOCS 3953 (the error bars for each flux measurement are within their markers). The SED of the BL (5600 K) candidate is plotted with the red shaded region while the SED of the potential WD companion (12,000 K) is represented with the blue shaded region. The combined flux from the BL and WD companion is plotted with a black line. The green triangle represents the $3\sigma$ upper limit on a GALEX $FUV$ flux measurement. The lower plot shows the residuals between the BL model spectrum and flux measurements.}
    \label{fig:sed}
\end{figure*}

WOCS 3953 (Gaia DR3~573956069112683392) presents an intriguing case as a BL candidate in NGC 188. BLs were first identified and characterized in M67 by \cite{leiner_blue_2019} as main-sequence stars with anomalously rapid rotation periods (2-8 days compared to the 20-30 days periods expected at that age). These stars have much in common with BSSs, including a number of long-period companions. (BLs identified by rapid rotation do not have close companions by selection.) In one case, a UV excess revealed a hot white dwarf (WD) companion \citep{nine_wiyn_2023, leiner_blue_2025}, providing strong evidence that these are mass-transfer products. Crucially, BLs differ from BSSs in that they photometrically blend into the main-sequence population. Nonetheless, they share many physical properties with BSSs---including binary mass-transfer signatures such as WD companions, elevated rotation, and long-period, low-to-modest eccentricity orbits---suggesting that they may represent a lower-luminosity extension of the same formation pathways \citep{mathieu_blue_2025}. 

 The system is a binary with an orbital period of 940 days and an eccentricity measured at $0.04 \pm 0.04$ (Table~\ref{table:orbits}), consistent with a circular orbit. Circular orbits for main-sequence binaries of such long periods are rare, and suggest prior tidal or other dissipative processes, for example, during RLOF mass transfer from an earlier evolved companion. Although observational evidence indicates that RLOF will not always circularize orbits, circular orbits at such long periods are seen in some mass-transfer products \citep{mathieu_blue_2025}. As shown in Figure~\ref{fig:3953_cmd}, WOCS 3953 lies at the blue edge of the NGC 188 main sequence. Although BLs are often embedded within the main sequence, this slight photometric offset may indicate modest mass gain, placing WOCS 3953 near the boundary between BLs and BSSs.

To further investigate this star, we compiled a comprehensive spectral energy distribution (SED) from 77 photometric flux measurements across a wide wavelength range, including data from Spitzer \citep{peter_spitzer_2019}, WISE \citep{schlafly_unwise_2019}, 2MASS \citep{zacharias_naval_2004}, Pan-STARRS \citep{chambers_pan-starrs1_2019}, Gaia \citep{vallenari_gaia_2023}, WIYN (\citealt{sarajedini_wiyn_1999}, \citetalias{gellerWIYNOpenCluster2008}), GALEX \citep{bianchi_revised_2017}, and SWIFT \citep{yershov_serendipitous_2014}, as shown in Figure~\ref{fig:sed}. Following the procedure described in \citet{nine_wiyn_2023}, we modeled the primary star (i.e., the BL) SED with \citet{castelli_new_2003} stellar atmosphere models, normalized to the observed $V$-magnitude. Adopting the cluster metallicity from our isochrone fit, we perform a $\chi^2$-minimization (using flux measurements $> 3000 \; \mathrm{\mathring{A}}$) to find the best fit main-sequence model and use a Monte Carlo simulation to estimate uncertainties. We find that the best-fit parameters for the primary star are $T_{\mathrm{eff}} = 5600 \pm 25$ K and $\log{g} = 4.5 \pm 0.1$ cm s$^{-2}$ (Figure~\ref{fig:sed}), corresponding to a radius of $1.14  \pm 0.003\; {R}_\odot$ at the Gaia-derived distance for NGC 188 ($1820 \pm 5$ pc).

A notable ultraviolet excess is apparent around $\mathrm{1930\; \mathring{A}}$ ($UVW2$) in measurements from the SWIFT UVOT catalog \citep{yershov_serendipitous_2014}. UV excesses at these wavelengths can be indicative of a hot WD companion, as frequently observed in post-mass-transfer systems \citep[e.g.,][]{gosnell_implications_2015, nine_wiyn_2023, rani_uocs_2021}. Chromospheric UV excesses of this magnitude are unlikely at the primary effective temperature. 

We further investigate the possible presence of a WD by including \citet{koester_white_2010} WD model atmospheres in our SED fits. As the SWIFT $UVW2$ excess at $\mathrm{1930\; \mathring{A}}$ is just one photometric point in the UV, it does not provide sufficient leverage to constrain the temperature of our WD models; an additional far-UV measurement is needed. Although this star has a reliable $NUV$ detection from GALEX, \cite{bianchi_revised_2017} flagged the $FUV$ measurement as being located on a detector hotspot and reported no detection. This allows us to place an FUV flux upper limit on the hottest WD that would not have been detected in the GALEX $FUV$ image of this field. Using the same pointing that imaged the location of WOCS 3953 in the far-UV \citep{bianchi_revised_2017}, we take our upper limit of detection as the $3\sigma$ upper limit from the faintest GALEX $FUV$ source on that image ($22.8 \pm 0.55$ mag) to account for the unknown contribution from the hotspot. We then normalize the combined flux from the BL and WD to the observed $UVW2$ and find that the upper limit on a WD companion temperature of 12,000 K and $\log{g} = 7.5$ cm s$^{-2}$ (Figure~\ref{fig:sed}), corresponding to a radius of $4 \pm 0.3\; {R}_\oplus$ at the Gaia-derived distance of the cluster. Even so, we find that our model is unable to fully reconcile the measured fluxes in the near-UV (as seen in the inset of Figure~\ref{fig:sed}). Further UV observations, are required to definitively confirm the presence of a WD companion. 

Using our derived primary mass from the cluster isochrone fit of $m_1 = 1.03 \; {M_\odot}$ and the binary mass function, we calculate a minimum companion mass of $\sim 0.2 \; {M_\odot}$, permitting a He or CO WD from red giant or asymptotic giant branch mass transfer, respectively.

WIYN Hydra spectroscopy reveals no sign of rapid rotation in WOCS 3953. The observed CCF profile is narrow, corresponding to a rotational velocity below our detection limit of 10 $\mathrm{km\;s^{-1}}$ on $v \sin{i}$. Although BLs identified in M67 typically exhibit rapid rotation linked to recent mass transfer, the absence of rapid rotation in WOCS 3953 does not preclude a past mass-transfer event as it may have happened long enough ago for the BL to have spun down \citep{nine_wiyn_2023}. 

Finally, a prior WOCS photometric study by \citet{kafka_wiyn_2003} suggested WOCS 3953 as a possible short-period (0.88-day) eclipsing binary. However, subsequent reassessment by the authors (S. Kafka, private communication) revealed this classification to be unreliable. Using 364 photometric measurements from the Zwicky Transient Facility \citep[ZTF;][]{masci_zwicky_2019}, we find that the observed fluxes exhibit variability at the $\sim 1.8 \sigma$ level compared to their reported uncertainties. A Lomb-Scargle periodogram did not find a significant signal, albeit noting that the sampling was every 2 to 4 days. New targeted high-precision time-series photometry of WOCS 3953 may be informative.

Despite having some properties of a star that has undergone mass transfer, we note this star as a \textit{candidate} BL given our current inability to either confirm or reject its mass transfer origin from having a hot WD companion.

\subsubsection{WOCS 4945: A Faint Blue Straggler Candidate}
\label{subsec:wocs4945}

WOCS 4945 (Gaia DR3~573943046772016768) is both the faintest BSS (roughly 1.5 magnitude below MSTO; refer to Section~\ref{subsec: membership_changes} for more information on this star's membership) and one of the longest-period binaries detected in NGC 188 ($P = 5510.0 \pm 50$ days, $e=0.48 \pm 0.07$). This star is offset in the CMD to the blue of the ZAMS, indicating mass gain. 

Using our isochrone parameters, we fit the CMD location of WOCS 4945 with \textsf{MIST} evolutionary tracks and determined the star to have a photometric mass of $0.98\pm 0.04 \; M_\odot$. The minimum mass of the companion is $0.75\pm 0.02\; M_\odot$, giving a mass ratio ($q = \frac{m_{\rm companion}}{m_{\rm BSS}}$) of at least 0.76. Stars of $q \gtrapprox 0.7$ tend to appear as double-lined spectroscopic binaries in WOCS observations, but none of the spectra or cross-correlations show evidence of a bright main-sequence companion, suggesting that the companion is a WD star. If this system is a result of mass transfer, a WD of this mass would require the progenitor to have been an asymptotic giant branch star of approximately 3 $M_\odot$. For NGC 188, current asymptotic giant branch masses are $\sim1.2 \, M_\odot$, so either the interaction that formed this system occurred many Gyr ago (as the lifespan of a $3\; M_\odot$ star is several hundred Myr) or the progenitor donor was itself a BSS, potentially formed through the merger of two main sequence stars \citep[e.g.,][]{leiner_blue_2025, shariat_once_2024}. Alternatively, if the companion is not a WD, but a K-dwarf, that would suggest the current BSS is the result of a merger of two low-mass stars.

Fitting an SED to WOCS 4945 following the procedure described for WOCS 3953 yields an ultraviolet excess from SWIFT $UVW2$ measurements. Like with WOCS 3953, GALEX reports a low $NUV$ mag (21.83 mag) and no $FUV$ measurement \citep{bianchi_revised_2017}. Given the lower SWIFT $UVW2$ flux measurement for this star, the upper limit on the GALEX $FUV$ flux does not provide a meaningful constraint on a possible WD companion temperature. Additional far-UV measurements are necessary to put constraints on a possible WD companion. 

Eccentric, long-period post-mass transfer binary systems are common (see for example Figure 4 of \cite{mathieu_blue_2025}). These systems may provide key insights into unanswered questions about the conditions under which stable mass transfer between BSS progenitors can occur. \cite{sepinsky_interacting_2007} found that $q_{\rm crit}$ for eccentric orbits to widen during conservative mass transfer to be strongly dependent on the eccentricity of the system. Equation 41 from that work gives $q_{\rm crit} = 0.85$ for the current eccentricity of WOCS 4945, suggesting that the orbit of WOCS 4945 was expanding at the end of mass transfer, which may have created a stabilizing effect on mass transfer. Using the $\delta$-function mass transfer models of \cite{sepinsky_interacting_2009} that account for non-conservative mass transfer, \cite{rocha_mass_2025} modeled mass transfer in eccentric high-mass binary systems with detailed stellar evolution, and found that systems with similar characteristics to WOCS 4945 ($10^3 < P < 10^4$ days and $q_{initial} > 1.3$, although of very different masses) remained eccentric after mass transfer provided they had a large initial eccentricity. They further found that eccentric systems underwent stable mass transfer (instead of undergoing unstable mass transfer and a common envelope) in 80\% of the cases they investigated. 

Both WOCS 4945 and WOCS 3953 appear to be post-interaction binary stars of almost identical BSS masses and periods $\gtrapprox 1000$ days, but the former is in an eccentric orbit and the latter is in a circular orbit. The periastron separation of WOCS 4945 is $\sim8$ times greater than that of WOCS 3953 (barring factors of $\sin{i}$ for both). Interestingly, the WOCS 4945 CMD location is bluer than the ZAMS whereas WOCS 3953 is still embedded in the main sequence, perhaps indicative of different amounts of mass gained. Neither BSS shows spectroscopic evidence of rapid rotation. 

At the time of Hubble Space Telescope far-UV observations of the NGC 188 BSSs by \cite{gosnell_detection_2014}, WOCS 4945 and WOCS 3953 were not yet identified as mass transfer candidates so neither were observed for possible WD companions. Future UV observations of this cluster should seek to determine temperature and $\log g$ of their possible WD companions, as these could constrain the progenitors of these similar luminosity BSSs. With formation histories of these stars, their similar final masses but somewhat different periastron separations may help inform the impacts of tides during binary interactions and explain why WOCS 3953 circularized whereas WOCS 4945 did not.

\subsubsection{WOCS 4230: A Close Binary Blue Straggler with a WD Companion}

WOCS 4230 is a BSS with a circular period of 0.456 days. This star also has measured Hubble Space Telescope far-UV flux (F165LP, F150LP, F140LP, F150N, and F140N) indicative of a $T_{\rm eff} = 11800 \pm 500$ WD companion \citep{gosnell_implications_2015}.

WOCS 4230 is photometrically variable with sinusoidal variations \citep[$P=0.3102$ days][]{popov_new_2013}. \cite{song_variable_2023} found it to have a period of 0.227(1) days, exactly half of the orbital period we report here, suggesting the photometric period to be a harmonic of the true period. If the photometric period is the same as the orbital period, then WOCS 4230 is tidally synchronized.

We found WOCS 4230 to have $v\sin i= 106\rm\;km\;s^{-1}$. We do not report formal errors for our $v\sin i$ measurements as they are based on the median full-width half-max of our CCFs, but given that this measurement is near the maximum to which we are sensitive, we estimate the error to be at least $10\rm\;km\;s^{-1}$, which we adopt for the discussion below. 

Using our isochrone parameters, we fit the CMD location of WOCS 4230 with \textsf{MIST} evolutionary tracks and determined the star to have a photometric mass of $1.20\pm 0.02 \; M_\odot$---giving the companion a minimum mass of $0.19\pm 0.01\;  M_\odot$---and radius of $1.39\pm 0.06\; R_\odot$. If the star is tidally synchronized with the orbital period, the star has a rotation rate of $154\pm6.6\rm\;km\;s^{-1}$, approximately one-third the break-up speed of a star of this mass and radius. Comparing this against our $v\sin i$ measurement gives $i = 43.6\pm5.6$, which in turn suggests $m_2 = 0.30\pm 0.04\;  M_\odot$, meaning the companion would be a He WD. From \textsf{MIST} evolutionary tracks, this mass corresponds to the core mass of a giant donor of $m \sim1.2\; M_\odot$ and $r \sim25\pm15\; R_\odot$, which would mean the interaction occurred when the donor was on the lower portion of the red giant branch and had an initial orbital period between $10^1-10^2$ days (see Section~\ref{subsec:period_phase}).

As the orbital period of this system is 0.4556 days, it falls about two orders of magnitude below the WD mass-final orbital period relationship for stable mass transfer of \cite{rappaport_relation_1995}, suggesting that this system underwent a period of unstable mass transfer during a common envelope event. Such an event is not expected to transfer mass to the companion \citep{ivanova_common_2011}. However, the CMD mass of WOCS 4230 ($1.20\pm 0.02 \;\rm M_\odot$) is larger than the turn-off mass of the cluster ($\sim1.1\;\rm M_\odot$), which given the relatively recent WD cooling age \citep[$350 \;\rm Gyr$][]{gosnell_implications_2015} suggests that the BSS gained some mass during the interaction. If mass transfer began on the subgiant branch, the system could have undergone stable mass transfer \citep{petrovic_influence_2021,leiner_origin_2017}, potentially before a common envelope developed as the donor climbed the red giant branch and rapidly expanded.

\subsubsection{WOCS 5020: The Longest Period Blue Straggler Star}
\label{subsec:wocs5020}

With a preliminary orbit solution having a period of $38,000$ days, WOCS 5020 has the longest known period of a cluster BSS of which we are aware, although longer-period BSSs have been found in the field \citep[e.g.,][]{danilovich_chemical_2024}. The BSS has a CMD mass of $1.52 \pm 0.02 \; M_\odot$, resulting in a minimum secondary mass of $0.55 \pm 0.12 \;M_\odot$. Even acknowledging the large error on the secondary mass driven by uncertainties in the orbit, the minimum companion mass is consistent with the mass of a CO WD formed from a progenitor with the same initial mass as the giants in NGC 188. Our orbital solution and CMD masses suggests the system has a current minimum periastron separation ($a \sin i \cdot (1-e)$) of 7.3 AU, which may be beyond the regime of the asymptotic giant branch donor's envelope filling its Roche lobe but well-within the regime of wind RLOF \citep{sun_wind_2024}. Further, the eccentricity of the system can change dramatically over the course of mass transfer \citep{rocha_mass_2025}, so these stars may not have always had the same separation. 

We note that we have observed significantly less than one orbital cycle (as P$\simeq100$ years), so it is possible that these data could be fit by an even longer period and more eccentric orbit.

The first RV measurement ($-45.7$ km $s^{-1}$) alone drives the reported center-of-mass RV, which is somewhat low for the cluster. All of the data from WIYN suggest that this star is a RV member of the cluster.

Formally, WOCS 5020 ($e/i = 3.9$) falls below our threshold for velocity variability. Indeed, if we were to have observed its RVs for a 10,000 day range starting now (roughly orbital phase of 0.2), our orbit suggests our measurements would systematically decrease by only about 1 km $s^{-1}$, far below our velocity-variable threshold. This might provide insight into the origin of some of the apparent single star BSSs. For example, WOCS 4290, a narrow-lined apparently single BSS ($e/i = 1.04$), shows RVs that have slowly increased by about 1 km $s^{-1}$ over 10,000 days of observations on WIYN.

\section{Summary} \label{sec:summary}
This study continues the WOCS long-term spectroscopic monitoring of the old open cluster NGC 188. By combining new observations with archival data and updated astrometric constraints, we re-examine this cluster's binary population. 
\begin{enumerate}
    \item \textbf{RV Database}: Our RV dataset spans nearly three decades and includes over 10,500 measurements. This extended baseline increases our sensitivity to long-period binaries and refines orbital solutions for previously identified systems. 
    \item \textbf{Cluster Membership and Properties}: Using Gaia DR3 proper-motions and parallaxes, we determine a membership sample of 546 stars within $0.5^\circ$ of the cluster center and G $ \leq 16.5$. Isochrone fitting with \textsf{MIST} models yields an age of $6.4 \pm 0.2$ Gyr and a distance of $1820 \pm 5$ pc for NGC 188. 
    \item \textbf{Binary Statistics}: We report 35 new single-lined spectroscopic binary orbits. The observed main-sequence spectroscopic binary frequency is $24.5 \pm 2.8\%$. After incompleteness corrections for undetected binaries, this yields a binary frequency of $33.1 \pm 3.8\%$ for $P < 10^4$ days.
    \item \textbf{Tidal Circularization}: We use a MCMC-based approach for the analysis of the period-eccentricity distribution of main-sequence binaries. Our methodology gives a tidal circularization period $14.4^{+0.14}_{-0.11}$ days, consistent with previous measurements but determined here with improved precision, given our larger sample of binary orbits. 
    \item \textbf{Giant Binary Evolution}: We find that binaries with evolved primary stars show a lower envelope of orbital periods that rises with decreasing $\log g$. Compared to the main-sequence period distribution, giants are systematically missing at small separations---consistent with mass transfer or common-envelope evolution converting these systems into BSSs and BLs.
    \item \textbf{Mass Transfer Candidates}: We identify eighteen secure BSSs, four BSS candidates, and one newly identified yellow straggler, of which 20 (87\%) are in binaries. Among the stars of note, we discuss four stars: WOCS 3953 is a BL candidate with a long-period circular orbit and ultraviolet excess; WOCS 4230 has a 0.456-day period with a very close WD companion; WOCS 4945 is a BSS $\sim1.5$ mag fainter than the MSTO with a companion more massive than a CO WD that current giants of NGC 188 could produce; and WOCS 5020 is the longest-period cluster BSS yet detected.
    
\end{enumerate}

\begin{acknowledgements}
The authors express their gratitude to the staff of the WIYN Observatory, without whom we would not have been able to collect these thousands of stellar spectra. We thank the many undergraduate and graduate students who have obtained spectra over the years at WIYN for this project. R.S.N. would like to thank Don Dixon and Andrew Nine for their support with SED fitting techniques and Bob Aloisi for his guidance on time-series photometry. Finally, we acknowledge the support of the National Science Foundation through award NSF AST-1714506, the Wisconsin Space Grant Consortium through awards RFP22\_10-0, RFP25\_4-0, and RFP25\_12-0, the Wisconsin Alumni Research Fund, and the University of Wisconsin-Madison L\&S Honors Program for providing support through an Honors Summer Research Apprenticeship. For our computational needs, our pipeline used \textsf{HTCondor}, a high-throughput computing cluster offered by the Center for High Throughput Computing (CHTC) at the University of Wisconsin-Madison. 

This work has made use of data from the European Space Agency (ESA) mission Gaia (\url{https://www.cosmos.esa.int/gaia}), processed by the Gaia Data Processing and Analysis Consortium (DPAC; \url{https://www.cosmos.esa.int/web/gaia/dpac/ consortium}). Funding for the DPAC has been provided by national institutions, in particular the institutions participating in the Gaia Multilateral Agreement. 

This work was conducted at the University of Wisconsin-Madison, which is located on occupied ancestral land of the Ho-Chunk people, a place their nation has called Teejop since time immemorial. In an 1832 treaty, the Ho-Chunk were forced to cede this territory. The university was founded on and funded through this seized land; this legacy enabled the science presented here. Observations for this work were conducted at the WIYN telescope on Kitt Peak, which is part of the lands of the Tohono O’odham Nation.
\end{acknowledgements}

\facilities{WIYN - Wisconsin-Indiana-Yale-NOAO Telescope (Hydra MOS), Gaia, SDSS, GALEX, Pan-Starrs, 2MASS, Spitzer, SWIFT, WISE.}

\software{\textsf{Astropy} \citep{astropy_collaboration_astropy_2013,astropy_collaboration_astropy_2022,almendros-abad_spectroscopic_2023}, \textsf{MIST} \citep{dotter_mesa_2016,choi_mesa_2016,paxton_modules_2011,paxton_modules_2013,paxton_modules_2015}, \textsf{NumPy} \citep{harris_array_2020}, \textsf{pandas} \citep{mckinney_data_2010}, \textsf{SciPy} \citep{virtanen_scipy_2020}, \textsf{scikit-learn} \citep{pedregosa_scikit-learn_2011}, \textsf{pysynphot} \citep{stsci_development_team_pysynphot_2013}, \textsf{The Joker} \citep{price-whelan_joker_2017}, \textsf{emcee} \citep{foreman-mackey_emcee_2013}, \textsf{glue} \citep{beaumont_hackable_2015,robitaille_glueviz_2019}, \textsf{HTCondor} \citep{thain_distributed_2005}}, \textsf{dustmaps} \citep{green_dustmaps_2018}. 

\appendix \label{sec:appendix}

\startlongtable
\begin{deluxetable*}{cccccccccccc} \label{table:field_orbits}
\tablewidth{0pt}
\tabletypesize{\scriptsize}
\tablecaption{Orbital Parameters For Field Single-Lined Binaries}

\tablehead{
   \colhead{ID} & \colhead{P} & \colhead{Orbital} & \colhead{$\gamma$} & \colhead{K} & \colhead{$ecc$} & \colhead{$\omega$} & \colhead{T$_0$} & \colhead{$a \; \sin(i)$} & \colhead{$f(m)$} & \colhead{$\sigma$} & \colhead{N} \\
   \colhead{} & \colhead{(days)} & \colhead{Cycles} & \colhead{(km s$^{-1}$)} & \colhead{(km s$^{-1}$)} & \colhead{} & \colhead{(deg)} & \colhead{(HJD-2,400,000 d)} & \colhead{($10^6$ km)} & \colhead{($M_\odot$)} & \colhead{(km s$^{-1}$)} & \colhead{}
}

\startdata
160 & 306.25 & 15.9 & -36.11 & 7.24 & 0.295 & 330. & 57468.0 & 29.1 & $1.05*10^{-2}$ & 0.259 & 16 \\ 
 & $\pm$ 0.21 &  & $\pm$ 0.08 & $\pm$ 0.10 & $\pm$ 0.016 & $\pm$ 3 & $\pm$ 2.6 & $\pm$ 0.4 & $\pm$ $4.0*10^{-4}$ &  &  \\ 
200 & 3030 & 1.2 & -65.2 & 8.4 & 0.41 & 86 & 54240 & 319 & $1.4*10^{-1}$ & 1.262 & 15 \\ 
 & $\pm$ 110 &  & $\pm$ 0.5 & $\pm$ 0.6 & $\pm$ 0.05 & $\pm$ 15 & $\pm$ 70 & $\pm$ 23 & $\pm$ $3.0*10^{-2}$ &  &  \\ 
216 & 286.38 & 19.6 & -80.7 & 12.5 & 0.06 & 50 & 56110 & 49 & $5.7*10^{-2}$ & 0.393 & 10 \\ 
 & $\pm$ 0.22 &  & $\pm$ 0.4 & $\pm$ 1.2 & $\pm$ 0.03 & $\pm$ 80 & $\pm$ 70 & $\pm$ 5 & $\pm$ $1.7*10^{-2}$ &  &  \\ 
398 & 133.15 & 21.7 & -85.80 & 12.7 & 0.562 & 180. & 54998.7 & 19.2 & $1.6*10^{-2}$ & 0.385 & 9 \\ 
 & $\pm$ 0.24 &  & $\pm$ 0.22 & $\pm$ 1 & $\pm$ 0.022 & $\pm$ 4 & $\pm$ 1.6 & $\pm$ 1.5 & $\pm$ $4.0*10^{-3}$ &  &  \\ 
1178 & 1894 & 3.0 & -39.83 & 8.18 & 0.42 & 232 & 54891 & 193 & $8.0*10^{-2}$ & 0.332 & 13 \\ 
 & $\pm$ 12 &  & $\pm$ 0.12 & $\pm$ 0.15 & $\pm$ 0.03 & $\pm$ 3 & $\pm$ 10. & $\pm$ 4 & $\pm$ $5.0*10^{-3}$ &  &  \\ 
1888 & 2207 & 4.4 & -41.51 & 6.6 & 0.24 & 359 & 52470 & 195 & $6.0*10^{-2}$ & 1.025 & 44 \\ 
 & $\pm$ 11 &  & $\pm$ 0.16 & $\pm$ 0.3 & $\pm$ 0.03 & $\pm$ 10. & $\pm$ 60 & $\pm$ 8 & $\pm$ $8.0*10^{-3}$ &  &  \\ 
1988 & 91.180 & 68.5 & -30.57 & 6.98 & 0.046 & 318 & 54796 & 8.75 & $3.21*10^{-3}$ & 0.309 & 11 \\ 
 & $\pm$ 0.025 &  & $\pm$ 0.11 & $\pm$ 0.14 & $\pm$ 0.021 & $\pm$ 25 & $\pm$ 6 & $\pm$ 0.18 & $\pm$ $1.9*10^{-4}$ &  &  \\ 
2225 & 1.60533 & 1802.2 & -41.81 & 3.0 & 0.22 & 160 & 55085.4 & 0.064 & $4.0*10^{-6}$ & 0.398 & 10 \\ 
 & $\pm$ 0.00003 &  & $\pm$ 0.25 & $\pm$ 1.2 & $\pm$ 0.16 & $\pm$ 60 & $\pm$ 0.3 & $\pm$ 0.025 & $\pm$ $5.0*10^{-6}$ &  &  \\ 
2442 & 282.05 & 18.9 & -26.86 & 4.20 & 0.340 & 2 & 56754 & 15.3 & $1.80*10^{-3}$ & 0.131 & 13 \\ 
 & $\pm$ 0.20 &  & $\pm$ 0.08 & $\pm$ 0.08 & $\pm$ 0.018 & $\pm$ 6 & $\pm$ 5 & $\pm$ 0.3 & $\pm$ $1.1*10^{-4}$ &  &  \\ 
2493 & 2.8544 & 1903.8 & 9.8 & 16.8 & 0.03 & 100 & 55195.8 & 0.66 & $1.4*10^{-3}$ & 1.347 & 10 \\ 
 & $\pm$ 0.0003 &  & $\pm$ 0.9 & $\pm$ 2.2 & $\pm$ 0.05 & $\pm$ 300 & $\pm$ 2.2 & $\pm$ 0.09 & $\pm$ $5.0*10^{-4}$ &  &  \\ 
2642 & 941 & 6.2 & -85.51 & 4.7 & 0.05 & 230 & 54590 & 61 & $1.03*10^{-2}$ & 0.576 & 17 \\ 
 & $\pm$ 6 &  & $\pm$ 0.17 & $\pm$ 0.3 & $\pm$ 0.05 & $\pm$ 60 & $\pm$ 160 & $\pm$ 3 & $\pm$ $1.7*10^{-3}$ &  &  \\ 
2815 & 751.9 & 7.5 & 24.67 & 4.3 & 0.25 & 33 & 56510 & 43 & $5.7*10^{-3}$ & 0.524 & 11 \\ 
 & $\pm$ 1.9 &  & $\pm$ 0.18 & $\pm$ 0.4 & $\pm$ 0.07 & $\pm$ 15 & $\pm$ 30 & $\pm$ 4 & $\pm$ $1.5*10^{-3}$ &  &  \\ 
3086 & 1368 & 4.8 & -71.20 & 4.80 & 0.067 & 185 & 57600 & 90.2 & $1.56*10^{-2}$ & 0.425 & 50 \\ 
 & $\pm$ 4 &  & $\pm$ 0.07 & $\pm$ 0.10 & $\pm$ 0.020 & $\pm$ 16 & $\pm$ 60 & $\pm$ 1.9 & $\pm$ $1.0*10^{-3}$ &  &  \\ 
3259 & 102.05 & 60.4 & -46.1 & 9.2 & 0.21 & 264 & 55564 & 12.6 & $7.7*10^{-3}$ & 1.0 & 21 \\ 
 & $\pm$ 0.04 &  & $\pm$ 0.3 & $\pm$ 0.5 & $\pm$ 0.04 & $\pm$ 15 & $\pm$ 4 & $\pm$ 0.6 & $\pm$ $1.2*10^{-3}$ &  &  \\ 
3623 & 495.1 & 10.8 & 17.00 & 11.8 & 0.48 & 284 & 56079 & 71 & $5.7*10^{-2}$ & 0.676 & 13 \\ 
 & $\pm$ 0.8 &  & $\pm$ 0.26 & $\pm$ 0.5 & $\pm$ 0.04 & $\pm$ 6 & $\pm$ 5 & $\pm$ 4 & $\pm$ $8.0*10^{-3}$ &  &  \\ 
3693 & 1290.7 & 6.3 & -25.58 & 2.72 & 0.470 & 357 & 56793 & 42.5 & $1.84*10^{-3}$ & 0.149 & 21 \\ 
 & $\pm$ 1.5 &  & $\pm$ 0.04 & $\pm$ 0.05 & $\pm$ 0.014 & $\pm$ 3 & $\pm$ 10. & $\pm$ 0.9 & $\pm$ $1.1*10^{-4}$ &  &  \\ 
3832 & 1410 & 4.6 & -42.4 & 9.6 & 0.41 & 347 & 56220 & 170. & $1.0*10^{-1}$ & 2.808 & 16 \\ 
 & $\pm$ 30 &  & $\pm$ 1.2 & $\pm$ 1.1 & $\pm$ 0.13 & $\pm$ 20. & $\pm$ 60 & $\pm$ 22 & $\pm$ $3.0*10^{-2}$ &  &  \\ 
3855 & 11.65003 & 529.9 & 8.61 & 21.33 & 0.014 & 70 & 54947.7 & 3.42 & $1.17*10^{-2}$ & 0.613 & 15 \\ 
 & $\pm$ 0.00019 &  & $\pm$ 0.17 & $\pm$ 0.23 & $\pm$ 0.009 & $\pm$ 60 & $\pm$ 1.9 & $\pm$ 0.04 & $\pm$ $4.0*10^{-4}$ &  &  \\ 
4145 & 97.229 & 59.0 & -36.27 & 26.0 & 0.231 & 183 & 56732.3 & 33.8 & $1.63*10^{-1}$ & 0.614 & 11 \\ 
 & $\pm$ 0.006 &  & $\pm$ 0.22 & $\pm$ 0.3 & $\pm$ 0.011 & $\pm$ 6 & $\pm$ 1.7 & $\pm$ 0.4 & $\pm$ $6.0*10^{-3}$ &  &  \\ 
4524 & 1263.5 & 7.6 & -42.78 & 5.36 & 0.131 & 26 & 51521 & 92.4 & $1.97*10^{-2}$ & 0.156 & 39 \\ 
 & $\pm$ 0.7 &  & $\pm$ 0.04 & $\pm$ 0.05 & $\pm$ 0.009 & $\pm$ 4 & $\pm$ 13 & $\pm$ 0.9 & $\pm$ $6.0*10^{-4}$ &  &  \\ 
4672 & 1.0312940 & 5675.4 & -58.0 & 40.1 & 0.19 & 160. & 56761.310 & 0.56 & $6.5*10^{-3}$ & 1.446 & 9 \\ 
 & $\pm$ 0.0000020 &  & $\pm$ 1 & $\pm$ 2.2 & $\pm$ 0.03 & $\pm$ 6 & $\pm$ 0.016 & $\pm$ 0.03 & $\pm$ $1.1*10^{-3}$ &  &  \\ 
5258 & 1.4059160 & 6598.6 & -40.2 & 46.0 & 0.020 & 30 & 52284.84 & 0.890 & $1.42*10^{-2}$ & 2.092 & 30 \\ 
 & $\pm$ 0.0000020 &  & $\pm$ 0.4 & $\pm$ 0.6 & $\pm$ 0.013 & $\pm$ 30 & $\pm$ 0.13 & $\pm$ 0.011 & $\pm$ $5.0*10^{-4}$ &  &  \\ 
5397 & 4070 & 2.1 & -41.56 & 3.81 & 0.31 & 180. & 52050 & 203 & $2.01*10^{-2}$ & 0.471 & 28 \\ 
 & $\pm$ 40 &  & $\pm$ 0.10 & $\pm$ 0.14 & $\pm$ 0.04 & $\pm$ 7 & $\pm$ 70 & $\pm$ 8 & $\pm$ $2.2*10^{-3}$ &  &  \\ 
5470 & 4090 & 2.3 & -42.06 & 3.39 & 0.25 & 203 & 54480 & 184 & $1.50*10^{-2}$ & 0.611 & 35 \\ 
 & $\pm$ 60 &  & $\pm$ 0.17 & $\pm$ 0.17 & $\pm$ 0.06 & $\pm$ 14 & $\pm$ 130 & $\pm$ 10. & $\pm$ $2.2*10^{-3}$ &  &  \\ 
5629 & 4.08671 & 2227.2 & -19 & 70. & 0.07 & 120 & 51743.2 & 3.9 & $1.5*10^{-1}$ & 13.324 & 22 \\ 
 & $\pm$ 0.00009 &  & $\pm$ 4 & $\pm$ 6 & $\pm$ 0.07 & $\pm$ 50 & $\pm$ 0.6 & $\pm$ 0.3 & $\pm$ $4.0*10^{-2}$ &  &  \\ 
5848 & 1750 & 2.0 & -33.6 & 6.4 & 0.45 & 27 & 52830 & 138 & $3.4*10^{-2}$ & 1.033 & 17 \\ 
 & $\pm$ 50 &  & $\pm$ 0.3 & $\pm$ 0.6 & $\pm$ 0.08 & $\pm$ 8 & $\pm$ 60 & $\pm$ 14 & $\pm$ $1.0*10^{-2}$ &  &  \\ 
6088 & 1067.1 & 3.8 & -82.10 & 6.04 & 0.100 & 182 & 57711 & 88.3 & $2.41*10^{-2}$ & 0.201 & 38 \\ 
 & $\pm$ 1.4 &  & $\pm$ 0.04 & $\pm$ 0.06 & $\pm$ 0.011 & $\pm$ 5 & $\pm$ 16 & $\pm$ 0.8 & $\pm$ $7.0*10^{-4}$ &  &  \\ 
6125 & 46.306 & 140.4 & -39.15 & 5.36 & 0.28 & 180. & 56820.6 & 3.28 & $6.5*10^{-4}$ & 0.474 & 10 \\ 
 & $\pm$ 0.005 &  & $\pm$ 0.16 & $\pm$ 0.22 & $\pm$ 0.06 & $\pm$ 8 & $\pm$ 0.9 & $\pm$ 0.15 & $\pm$ $8.0*10^{-5}$ &  &  \\ 
6405 & 6.33857 & 839.2 & -41.1 & 48.4 & 0.1 & 179 & 55168.3 & 4.20 & $7.4*10^{-2}$ & 3.864 & 12 \\ 
 & $\pm$ 0.00016 &  & $\pm$ 1.4 & $\pm$ 2.5 & $\pm$ 0.06 & $\pm$ 19 & $\pm$ 0.3 & $\pm$ 0.21 & $\pm$ $1.1*10^{-2}$ &  &  \\ 
6879 & 692 & 4.2 & -70.9 & 13 & 0.68 & 171 & 55103 & 90 & $6.0*10^{-2}$ & 0.481 & 11 \\ 
 & $\pm$ 4 &  & $\pm$ 0.8 & $\pm$ 7 & $\pm$ 0.11 & $\pm$ 6 & $\pm$ 11 & $\pm$ 50 & $\pm$ $9.0*10^{-2}$ &  &  \\ 
6952 & 553.2 & 5.2 & -85.0 & 12.8 & 0.32 & 221 & 54449 & 92 & $1.0*10^{-1}$ & 0.709 & 11 \\ 
 & $\pm$ 1.7 &  & $\pm$ 0.4 & $\pm$ 1.2 & $\pm$ 0.06 & $\pm$ 6 & $\pm$ 8 & $\pm$ 9 & $\pm$ $3.0*10^{-2}$ &  &  \\ 
7169 & 307.8 & 18.2 & -3.7 & 17 & 0.49 & 180. & 56164 & 64 & $1.1*10^{-1}$ & 1.827 & 17 \\ 
 & $\pm$ 0.3 &  & $\pm$ 1.3 & $\pm$ 5 & $\pm$ 0.11 & $\pm$ 7 & $\pm$ 6 & $\pm$ 20. & $\pm$ $1.0*10^{-1}$ &  &  \\ 
7297 & 196.09 & 24.9 & -22.84 & 3.14 & 0.662 & 328.3 & 57393.8 & 6.35 & $2.66*10^{-4}$ & 0.194 & 40 \\ 
 & $\pm$ 0.03 &  & $\pm$ 0.03 & $\pm$ 0.07 & $\pm$ 0.011 & $\pm$ 1.7 & $\pm$ 0.4 & $\pm$ 0.16 & $\pm$ $1.8*10^{-5}$ &  &  \\ 
7489 & 1274.3 & 4.5 & -62.08 & 6.86 & 0.02 & 230 & 61160 & 120. & $4.2*10^{-2}$ & 0.379 & 11 \\ 
 & $\pm$ 2.4 &  & $\pm$ 0.15 & $\pm$ 0.21 & $\pm$ 0.03 & $\pm$ 70 & $\pm$ 240 & $\pm$ 4 & $\pm$ $4.0*10^{-3}$ &  &  \\ 
7782 & 5.32767 & 1608.8 & -46.6 & 76.3 & 0.02 & 80 & 53125.9 & 5.59 & $2.45*10^{-1}$ & 7.486 & 33 \\ 
 & $\pm$ 0.00006 &  & $\pm$ 1.4 & $\pm$ 1.9 & $\pm$ 0.03 & $\pm$ 70 & $\pm$ 1.1 & $\pm$ 0.14 & $\pm$ $1.8*10^{-2}$ &  &  \\ 
7796 & 13.46564 & 607.6 & 21.6 & 14.9 & 0.25 & 342 & 53474.8 & 2.67 & $4.2*10^{-3}$ & 0.632 & 12 \\ 
 & $\pm$ 0.00021 &  & $\pm$ 0.3 & $\pm$ 1 & $\pm$ 0.04 & $\pm$ 9 & $\pm$ 0.4 & $\pm$ 0.17 & $\pm$ $8.0*10^{-4}$ &  &  \\ 
7975 & 338.7 & 16.6 & -47.4 & 7.8 & 0.24 & 355 & 57296 & 35.4 & $1.54*10^{-2}$ & 1.617 & 46 \\ 
 & $\pm$ 0.5 &  & $\pm$ 0.3 & $\pm$ 0.4 & $\pm$ 0.05 & $\pm$ 12 & $\pm$ 11 & $\pm$ 2.0 & $\pm$ $2.5*10^{-3}$ &  &  \\ 
8182 & 1329 & 3.8 & 0.8 & 11.9 & 0.65 & 242.2 & 57638 & 166 & $1.03*10^{-1}$ & 0.315 & 12 \\ 
 & $\pm$ 7 &  & $\pm$ 0.3 & $\pm$ 0.3 & $\pm$ 0.03 & $\pm$ 2.4 & $\pm$ 15 & $\pm$ 7 & $\pm$ $9.0*10^{-3}$ &  &  \\ 
8328 & 10.23171 & 548.2 & -19.99 & 7.70 & 0.23 & 281 & 56616.42 & 1.05 & $4.5*10^{-4}$ & 0.286 & 11 \\ 
 & $\pm$ 0.00012 &  & $\pm$ 0.13 & $\pm$ 0.26 & $\pm$ 0.03 & $\pm$ 5 & $\pm$ 0.12 & $\pm$ 0.04 & $\pm$ $4.0*10^{-5}$ &  &  \\ 
8768 & 543.6 & 9.9 & -45.1 & 19 & 0.89 & 100 & 55465 & 60 & $4.0*10^{-2}$ & 0.642 & 15 \\ 
 & $\pm$ 1.4 &  & $\pm$ 1.5 & $\pm$ 21 & $\pm$ 0.19 & $\pm$ 40 & $\pm$ 8 & $\pm$ 90 & $\pm$ $1.3*10^{-1}$ &  &  \\ 
9222 & 757 & 4.9 & 19.1 & 15.5 & 0.31 & 99 & 54668 & 154 & $2.5*10^{-1}$ & 1.121 & 10 \\ 
 & $\pm$ 4 &  & $\pm$ 0.4 & $\pm$ 0.7 & $\pm$ 0.03 & $\pm$ 12 & $\pm$ 23 & $\pm$ 8 & $\pm$ $4.0*10^{-2}$ &  &  \\ 
9240 & 39.680 & 111.2 & -47.55 & 20.26 & 0.308 & 135.5 & 54878.15 & 10.52 & $2.94*10^{-2}$ & 0.477 & 11 \\ 
 & $\pm$ 0.008 &  & $\pm$ 0.17 & $\pm$ 0.24 & $\pm$ 0.016 & $\pm$ 2.2 & $\pm$ 0.19 & $\pm$ 0.14 & $\pm$ $1.1*10^{-3}$ &  &  \\ 
9426 & 1053.2 & 6.2 & -47.9 & 16 & 0.78 & 331 & 55354 & 150 & $1.1*10^{-1}$ & 0.801 & 17 \\ 
 & $\pm$ 1.4 &  & $\pm$ 0.6 & $\pm$ 12 & $\pm$ 0.16 & $\pm$ 9 & $\pm$ 6 & $\pm$ 120 & $\pm$ $2.6*10^{-1}$ &  &  \\ 
9438 & 22.6131 & 143.9 & 16.36 & 9.5 & 0.240 & 236 & 55266.4 & 2.86 & $1.83*10^{-3}$ & 0.355 & 10 \\ 
 & $\pm$ 0.0012 &  & $\pm$ 0.13 & $\pm$ 0.3 & $\pm$ 0.017 & $\pm$ 10. & $\pm$ 0.6 & $\pm$ 0.1 & $\pm$ $1.9*10^{-4}$ &  &  \\ 
\enddata
\end{deluxetable*}
\input{ngc188_field.sb1pictures}

\bibliography{ngc188references}{}
\bibliographystyle{aasjournalv7}
\end{document}